\documentclass[copyright,creativecommons,noncommercial]{eptcs}
\providecommand{\event}{NCL' 22} 
\usepackage{breakurl}             
\usepackage{underscore}           

\usepackage[utf8]{inputenc} 
\usepackage{amsmath}
\usepackage{afterpage}
\usepackage{enumerate}
\usepackage{version}
\usepackage{turnstile} 
\usepackage{proof}
\usepackage{amssymb}
\usepackage{MnSymbol}
\usepackage{amsmath,amsfonts}

\usepackage{amsthm}
\usepackage{hyperref}
\usepackage{stmaryrd}
\usepackage{yfonts}
\usepackage{color}
\usepackage{algorithmicx}
\usepackage[compatible]{algpseudocode}
\usepackage{algorithm}
\usepackage{mdframed}
\usepackage{thmtools, thm-restate}
\usepackage{booktabs}
\usepackage{pgfplots}
\usepackage{tikz,color}
\usetikzlibrary{arrows,fit}
\usepackage{booktabs} 
\usepackage{array} 
\usepackage{paralist} 
\usepackage{verbatim} 
\usepackage{subfig} 
\usepackage[nottoc,notlof,notlot]{tocbibind} 
\usepackage[titles,subfigure]{tocloft} 
\usepackage{times}
\usepackage{turnstile} 
\usepackage{proof}
\usepackage{amssymb}
\usepackage{amsmath}
\usepackage{amsthm}
\usepackage[all]{xypic} 
\usepackage{hyperref}
\usepackage{stmaryrd}
\usepackage{yfonts}
\usepackage{color}
\usepackage{thmtools}
\usepackage{algorithmicx}
\usepackage[compatible]{algpseudocode}
\usepackage{algorithm}
\usepackage{mdframed}
\usepackage{array}
\usepackage{booktabs}
\usepackage{ wasysym }
\usepackage[normalem]{ulem}
\usepackage{tikz}
\usepackage{graphicx} 
\usepackage{multirow}

\newcommand{\Identity}{\mbox{Id}}

\newcommand{\RP}{\textsf{R}}
\newcommand{\PP}{\textsf{P}}

\newcommand{\langFOfunc}{\mathcal{L}^{\mathcal{P}}_{\logicFO}}

\newcommand{\co}{\textsf{c}}

\newcommand{\langFOfree}{\langFOfunc(\vec{x})}

\newcommand{\implyint}{\Rightarrow}

\newcommand{\langFOL}{\mathcal{L}_{\logicFO}}

\newcommand{\logicInt}{\textsf{Int}}

\newcommand{\logicLC}{\textsf{LC}}

\newcommand{\langIL}{\mathcal{L}_{\textsf{Int}}}

\newcommand{\vectoni}{\vec{\pm}_j}

\newcommand{\Bool}{\mathbb{B}}

\newcommand{\langProtoConexP}{\langProtoConex^{\mathbb{B}}}

\newcommand{\id}{id}

\newcommand{\comc}{c}

\newcommand{\vecmult}{\vec{k}}

\newcommand{\landst}{\land}

\newcommand{\veest}{\vee}

\newcommand{\langProto}{\langage}

\newcommand{\signatureb}{t}

\newcommand{\equivLang}{\rightsquigarrow}

\newcommand{\signature}{\tonicitytrace}

\newcommand{\simproto}{Z}
\newcommand{\simprotob}{Z}

\newcommand{\terms}{\mathcal{L}_{\logicFO}^{\var\cons}}

\newcommand{\tonicitytrace}{s}
\newcommand{\mult}{k}

\newcommand{\Assign}{\assign(\M,\Conex)}
\newcommand{\Assignb}{\assign(\M',\Conex)}

\newcommand{\assign}{\vec{\w}}

\newcommand{\NANDun}{~|_{\mbox{\tiny{1}}}~}
\newcommand{\NANDdeux}{~|_{\mbox{\tiny{2}}}~}
\newcommand{\NANDtrois}{~|_{\mbox{\tiny{3}}}~}

\newcommand{\eqdef}{\triangleq}

\newcommand{\Rset}{\mathcal{R}}

\newcommand{\z}{z}

\newcommand{\connec}[4]{~(\sigma_{#1},#3,\signature_{#2})~}

\newcommand{\genconn}[4]{\connec{#1}{#2}{#3}{#4}}

\newcommand{\connecb}[4]{~(\tau_{#1},#3,\signatureb_{#2})~}

\newcommand{\genconnb}[4]{\connecb{#1}{#2}{#3}{#4}}

\newcommand{\cmmm}{\genconn{1}{1}{\direct}{2}}

\newcommand{\cmmmpmp}{\genconn{2}{2}{\conv}{2}}

\newcommand{\cmmmpmpb}{\genconn{3}{2}{\conv}{2}}

\newcommand{\cmmmb}{\genconn{4}{1}{\direct}{2}}

\newcommand{\cmmmmpp}{\genconn{5}{3}{\conv}{2}}

\newcommand{\cmmmmppb}{\genconn{6}{3}{\conv}{2}}

\newcommand{\cmpm}{\genconn{1}{5}{\direct}{2}}

\newcommand{\cmpmppp}{\genconn{2}{4}{\conv}{2}}

\newcommand{\cmpmpmm}{\genconn{3}{6}{\direct}{2}}

\newcommand{\cmpmpmmb}{\genconn{4}{6}{\direct}{2}}

\newcommand{\cmpmpppb}{\genconn{5}{4}{\conv}{2}}

\newcommand{\cmpmmpmb}{\genconn{6}{5}{\direct}{2}}

\newcommand{\cpmm}{\genconn{1}{6}{\direct}{2}}

\newcommand{\cpmmpmm}{\genconn{2}{6}{\direct}{2}}

\newcommand{\cpmmppp}{\genconn{3}{4}{\conv}{2}}

\newcommand{\cpmmmpm}{\genconn{4}{5}{\direct}{2}}

\newcommand{\cpmmmpmb}{\genconn{5}{5}{\direct}{2}}

\newcommand{\cpmmpppb}{\genconn{6}{4}{\conv}{2}}

\newcommand{\cppm}{\genconn{1}{7}{\direct}{2}}

\newcommand{\cppmppm}{\genconn{2}{7}{\direct}{2}}

\newcommand{\cppmppmb}{\genconn{3}{7}{\direct}{2}}

\newcommand{\cppmppmc}{\genconn{4}{7}{\direct}{2}}

\newcommand{\cppmppmd}{\genconn{5}{7}{\direct}{2}}

\newcommand{\cppmppme}{\genconn{6}{7}{\direct}{2}}

\newcommand{\cmmpp}{\genconnb{2}{2}{\conv}{1}}

\newcommand{\cppmmr}{\genconnb{2}{1}{\direct}{1}}

\newcommand{\pitchforkb}{\bowtie}

\newcommand{\classGaggleLogics}{\mathbb{L}_{\textsf{GGL}}}

\newcommand{\conn}{\star} 

\newcommand{\Con}{\mathbb{C}} 

\newcommand{\Conex}{\textsf{C}}

\newcommand{\Conexb}{\textsf{C}'}

\newcommand{\Conexo}{{\textsf{C}_0}}

\renewcommand{\phi}{\varphi}

\newcommand{\x}{x}
\newcommand{\y}{y}

\newcommand{\bb}{\pm}

\newcommand{\complem}[1]{-{#1}}

\newcommand{\direct}{+}

\newcommand{\conv}{-}

\newcommand{\classConex}{\classmodel_{\Conex}}

\newcommand{\langProtoConex}{\langage_{\Conex}}

\newcommand{\var}{\mathcal{V}}
\newcommand{\cons}{\mathcal{C}}
\newcommand{\pred}{\mathcal{P}}

\newcommand{\prop}{\mathbb{A}}

\newcommand{\fe}{\mbox{\AE}}

\newcommand{\agts}{\mathbb{I}}
\newcommand{\langML}{\langage_{\logicML}}

\newcommand{\logicFO}{\textsf{FOL}}

\newcommand{\logicML}{\textsf{ML}}

\newcommand{\Prop}{\mathbb{P}}

\newcommand{\w}{w}

\newcommand{\letterp}{p}

\newcommand{\exto}[1]{\llbracket{#1}\rrbracket}

\newcommand{\classmodel}{\mathcal{M}} 

\newcommand{\BRAun}{\Yleft_{\mbox{\tiny{1}}}}

\newcommand{\BRAtrois}{\Yleft_{\mbox{\tiny{3}}}}

\newcommand{\RAdeux}{\Yright_{\mbox{\tiny{2}}}}
\newcommand{\RAtrois}{\Yright_{\mbox{\tiny{3}}}}

\newcommand{\fissiondeux}{\varoplus_{\mbox{\tiny{2}}}}

\newcommand{\fissionun}{\varoplus_{\mbox{\tiny{1}}}}

\newcommand{\classConexE}{\mathcal{E}_{\Conex}}

\newcommand{\langage}{\mathcal{L}}

\newcommand{\classint}{\mathcal{E}_{\textsf{Int}}}

\newcommand{\valid}[1]{\models_{#1}}

\newcommand{\W}{W}

\newcommand{\classup}{\mathcal{E}}

\newcommand{\structures}{\mathcal{M}_{\logicFO}}
\newcommand{\classtruc}{\mathcal{E}_{\logicFO}}

\renewcommand{\phi}{\varphi}

\newcommand{\grammarsep}{\text{~~}\mid\text{~~}}
\newcommand{\grammardef}{\text{~~}::=\text{~~}}

\newcommand{\nec}{\Box}

\newcommand{\bmid}{\mid}

\newcommand{\Kripke}{\mathcal{K}}

\newcommand{\assignment}{s}

\newcommand{\R}{R}

\newcommand{\AtomProp}{\mathbb{A}}

\newcommand{\M}{M} 

\newcommand{\simul}{\rightarrow}

\renewcommand{\models}{~ {\footnotesize{\sdtstile{}{}}}~ }

\newcommand{\entails}[1]{~{\footnotesize{\sststile{}{{}}}}~ }

\newcommand{\xR}{w}
\newcommand{\yR}{v}
\newcommand{\zR}{u}

\newcommand{\langR}{\langage\left(\textsf{Sub}_{-}\right)}
\newcommand{\langRa}{\langage\left(\textsf{Sub}\right)}

\newcommand{\implyun}{\supset_{\mbox{\tiny{1}}}}

\newcommand{\implys}{\supset}

\newcommand{\fusions}{\otimes}
\newcommand{\fusion}{\otimes_{\mbox{\tiny{3}}}}

\newcommand{\implybiss}{\subset}
\newcommand{\implybis}{\subset_{\mbox{\tiny{2}}}}

\newcommand{\entailR}{~ {\footnotesize{\dststile{}{}}}~ }

\renewcommand{\models}{~ {\footnotesize{\sdtstile{}{}}}~ }

\declaretheorem[style=definition,qed=$\dashv$]{definition}

\declaretheorem[style=definition,qed=$\dashv$]{example}
\theoremstyle{plain}
\newtheorem{theorem}{Theorem}
\newtheorem{lemma}{Lemma}
\newtheorem{fact}{Fact}

\newtheorem{proposition}{Proposition}

\theoremstyle{remark}

\newtheorem{remark}{Remark}

\renewcommand{\vec}{\overline}

\newcommand{\ddist}{4.5pt}

\usetikzlibrary{automata,arrows,positioning,snakes,shapes}

\tikzset{every picture/.style={inner sep=1.5pt}}
\tikzstyle{t}=[isosceles triangle,draw,anchor=east,inner sep=0pt,rotate=90]
\tikzstyle{w}=[draw,fill,circle]
\tikzstyle{d}=[draw,circle,minimum size=2*\ddist,after node path={node[circle,fill] at (\tikzlastnode)  {}}]
\tikzstyle{wl}=[draw,fill,circle]
\tikzstyle{ws}=[draw,circle]

\title{A  van Benthem Theorem for Atomic and Molecular Logics}
\author{Guillaume Aucher
\institute{Univ Rennes, CNRS, IRISA, IRMAR\\ Rennes, France}
\institute{263, Avenue du G{\'e}n{\'e}ral Leclerc\\
	35042 Rennes Cedex, France}
\email{guillaume.aucher@univ-rennes1.fr}
}
\def\titlerunning{A  van Benthem Theorem for Atomic and Molecular Logics}
\def\authorrunning{G. Aucher}
\begin{document}
\maketitle


\begin{abstract}
After recalling the definitions of  atomic and molecular logics, we show how notions of bisimulation can be automatically defined from the truth conditions of the connectives of any of these logics. Then,  we prove a generalization of  van Benthem modal characterization theorem for  molecular logics.  Our molecular connectives should be uniform and contain all conjunctions and disjunctions. We use modal logic, the Lambek calculus and modal intuitionistic logic as  case study and   compare in particular  our work with  Olkhovikov's work.
\end{abstract}

\section{Introduction}

Modal bisimulation is the notion of invariance of modal logic: every formula of first-order logic  (\logicFO)  with a free variable whose truth value is always the same in two bisimilar
models is equivalent to the  translation  into \logicFO\ of a formula of modal logic. This is the core of the van Benthem characterization theorem. 
A wide variety of    non--classical logics have been introduced over the past decades:  modal logics, relevant logics,    Lambek calculi,  to name just a few. For each of these logics, one can define a notion of invariance and prove by adapting the van Benthem's  characterization theorem  that this notion of invariance   characterizes the given logic  as a fragment of \logicFO. A drawback of this logical pluralistic approach  is that this has to be done by hand on a case by case  basis for each non--classical logic. Each time the notion of invariance has to be found out and each time the proof of the van Benthem characterization theorem  has to be adapted for that specific notion of invariance. For example, a similar characterization theorem has been proved for  (modal) intuitionistic logic \cite{Olk17}, temporal logic \cite{KurRij97}, sabotage modal logic \cite{AucEtAl18},    the modal $\mu$--calculus \cite{JanWal96}, graded modal logic \cite{Rij00}. 
 This situation is obviously   problematic if one shares the ideal of ``universal logic'' \cite{Bez07}. Instead, we would prefer to obtain automatically from  the definitions of the connectives of a given  logic  a suitable definition  of bisimulation and its associated characterization theorem. This is what we are going to provide in this article for a wide range of non--classical logics, those molecular logics whose connectives are uniform, a notion introduced in that paper.  Atomic and molecular logics are  introduced in \cite{Auc22}. They behave as `normal forms' for logics since we show in \cite{Auc22}  that every non-classical logic such that the truth conditions of its connectives can be expressed in terms of  first-order formulas is as expressive as an atomic or molecular logic. 

\paragraph{Organization of the article.}
We  start in Sections \ref{section:2} and \ref{sec:2.2.1}  by  recalling first--order logics, modal logic, the Lambek calculus and modal intuitionistic logic. 
In Section \ref{section:4} we recall atomic and molecular logics. 
Then, in Section \ref{sec:6}, we will show how a suitable notion of bisimulation/invariance  can be defined automatically from the definition of the connectives of any atomic or molecular logic. Then, in Section \ref{sec:5}, we will generalize van Benthem modal characterization theorem to molecular logics whose connectives are uniform.  
 Finally, we discuss related work in Section \ref{sec:8}, in particular the work of Olkhovikov. 

\section{Classical  logics}

\label{section:2}

 
 \label{sec:2.3.1}
The set $\pred\eqdef\left\{\RP_1,\ldots,\RP_n,\ldots\right\}$ is a  set of  \emph{predicate symbols}  
of arity $\mult_1,\ldots,\mult_n,\ldots$ respectively 
(one of them can be the identity predicate $=$ of arity 2),  
$\var\eqdef\left\{v_1,\ldots, v_n,\ldots \right\}$ is a   set of \emph{variables} and $\cons\eqdef\left\{c_1,\ldots, c_n,\ldots\right\}$ is a set of \emph{constants}. Each of these sets can be finite or infinite.  $v_1,v_2,v_3,\ldots$ are the names of the variables and we use the expressions $x,x_1, x_2,\ldots,y,y_1,y_2,\ldots,z,z_1,z_2,\ldots$ to refer to arbitrary variables or constants, which can be for example $v_{42},v_5, c_{101},c_{21},\ldots$ 
%
%
\index{Language ! First-order}

The \emph{first-order language} $\langFOfunc$  is defined inductively by the following grammars in BNF: 
\begin{align*}
		\terms:~~ &t  \grammardef x  \grammarsep \co\\ 
		\langFOfunc:~~ &\phi  \grammardef 
		\RP t\ldots t
		\grammarsep\bot \grammarsep(\phi\rightarrow\phi)
		\grammarsep\forall x\phi 
	\end{align*}
	where 
	$x\in\var$, $\co\in\cons$, 
	$t\in\terms$ and $\RP\in\pred$. 
	Elements of $\terms$ are called \emph{terms} and elements of  $\langFOfunc$ are called \emph{first--order formulas}. Formulas of the form $\RP t_1\ldots t_\mult$ are called \emph{atomic formulas}.  
	\index{Term}  \index{Formula ! Atomic}
	\index{Variable ! Free}
	\index{Variable ! Bound} 
	If $\phi\in\langFOfunc$, the \emph{Boolean negation} of $\phi$, denoted $\neg\phi$, is defined  by the abbreviation  
	$\neg\phi\eqdef(\phi\rightarrow\bot)$. 
	 We also  use the  abbreviations $\top\eqdef\neg\bot$, $(\phi\vee\psi)\eqdef(\neg\phi\rightarrow\psi)$,  $(\phi\land\psi)\eqdef\neg(\neg\phi\vee\neg\psi)$ and $(\phi\leftrightarrow\psi)\eqdef(\phi\rightarrow\psi)\land(\psi\rightarrow\phi)$ as well as  the abbreviations $\exists x\phi\eqdef\neg\forall x\neg\phi$, $\forall x_1\ldots x_n \phi\eqdef\forall x_1\ldots\forall x_n\phi$, $\exists x_1\ldots x_n \phi\eqdef\exists x_1\ldots\exists x_n\phi$ and  $\forall\vec{x}\phi\eqdef\forall x_1\ldots x_n\phi$ if $\vec{x}=(x_1,\ldots,x_n)$ is a  tuple of variables. 
	
	
		

			Let $\phi\in\langFOfunc$. An occurrence of a variable $x$ in $\phi$ is  \emph{free} (in $\phi$)  if, and only if, $x$ is not within the scope of a quantifier of $\phi$. A variable which is not free (in $\phi$)  is \emph{bound} (in $\phi$). We say that a formula of $\langFOfunc$ is a \emph{sentence}  (or is \emph{closed}) when it contains no free variable. We denote by $\phi(x_1,\ldots,x_\mult)$ a formula of $\langFOfunc$ whose free variables or constants  coincide  \emph{exactly} with  $x_1,\ldots,x_\mult$. In doing so, we depart from the literature in which this notation means that the free variables of $\phi$ are \emph{included} in $\{x_1,\ldots,x_\mult\}$. Free variables may be used to bind elements of two different subformulas. For example, the formula $\RP y x\vee \RP' x z$ with free variables $x, y, z$ will be evaluated in a structure in such a way that $x$ will be assigned the same element of the domain in the two subformulas $\RP y x$ and $\RP' x z$.   
			
			We denote by $\langFOfree$  the fragment of $\langFOfunc$   whose formulas all contain at least one free variable or constant. 

	\label{sec:2.3.2}
	
	
		A  \emph{structure}   is a tuple $\M\eqdef\left(W,\left\{\R_1,\ldots, \R_n,\ldots,
		c_1,\ldots,c_n,\ldots\right\}\right)$ 
		where: 
		\begin{itemize}
			\item $W$ is a non-empty set called the \emph{domain}; 
			\item $\R_1,\ldots,\R_n,\ldots$ are relations over $W$ with the same arity as $\RP_1,\ldots,\RP_n,\ldots$ respectively;
			\item $c_1,\ldots,c_n,\ldots\in W$ are  elements of the domain called \emph{distinguished elements}.
		\end{itemize} \index{Domain} 
		
		
		An \emph{assignment} over $M$  is a mapping $\assignment:\var\cup\cons\rightarrow W$ such that for all $\co_i\in\cons$, $\assignment(\co_i)=c_i$. If $\assignment$ is an assignment,  $\assignment[x:=w]$ is the same assignment as  $\assignment$ except that the value of the variable $x\in\var$ is assigned to $w$.  
		A pair of structure and assignement $(M,\assignment)$ is called a \emph{pointed structure}. 
		The class  of all pointed structures  $(M,\assignment)$ is denoted $\structures$. 
	
		
		

		

		
		
			The \emph{satisfaction relation} $\valid\logicFO\subseteq \structures\times\langFOfunc$ is defined inductively as follows. 
			Below, we write $(M,\assignment)\models\phi$ for $((M,\assignment),\phi)\in\valid\logicFO$.   
				
				\[\begin{array}{l@{\quad}c@{\quad}l}
					(\M,\assignment)\models \bot  & & \mbox{never};\\
					(\M,\assignment)\models \RP_i t_1\ldots t_k
					& \mbox{ iff } & (\assignment(t_1),\ldots,\assignment(t_k))\in R_i;\\
					(\M,\assignment)\models (\phi\rightarrow\psi) &
					\mbox{ iff } & \mbox{if }(\M,\assignment)\models\phi \mbox{ then }  (\M,\assignment)\models\psi;\\
					(\M,\assignment)\models \forall x\phi
					& \mbox{ iff } & (\M,\assignment[x:=w])\models\phi\mbox{ for all } w\in W.
				\end{array}\]
				In the literature \cite{ChaKei98}, $(M,\assignment)\models\phi(x_1,\ldots,x_\mult)$ is  sometimes denoted $M\models\phi(x_1,\ldots,x_\mult)[\w_1,\ldots,\w_\mult]$,
				$M\models$ $\phi[\w_1/x_1,\ldots,\w_\mult/x_\mult]$ or simply $M\models\phi[\w_1,\ldots,\w_\mult]$, with $w_1=\assignment(x_1),\ldots,w_\mult=\assignment(x_\mult)$. 
				 In that case, we say that $(M,\assignment)$ makes $\phi$ \emph{true}. We say that the formula $\phi\in\langFOfunc$ is \emph{realized  in $\M$} 
			when  there is an assignment $\assignment$ such that   $(\M,\assignment)\models\phi$. 
			We depart from the literature by treating constants on a par with variables: the denotation of constants is usually not dealt with by means of assignments.  Two (pointed) structures are \emph{elementarily equivalent} when they  make true the same sentences.

		A triple  of the form  $\left(\langFOL,\classtruc,\valid\logicFO\right)$ is   called the \emph{first--order logic associated to $\langFOL$ and $\classtruc$}. 
		If $\langFOL=\langFOfunc$, the triple $\left(\langFOfunc,\classtruc,\valid\logicFO\right)$ is called  \emph{pure predicate logic (associated to $\classtruc$)},  if $\langFOL=\langFOfree$, the triple $\left(\langFOfree,\classtruc,\valid\logicFO\right)$ is called   \emph{pure predicate  logic with free variables and constants  (associated to $\classtruc$)}. 
		When $\classtruc$ is $\structures$, they are simply called respectively 
		\emph{pure predicate logic} and \emph{pure predicate  logic with free variables and constants}.

\section{Non-classical logics} 

\label{sec:2.2.1}

In this section, $\AtomProp$ is a  set of \emph{propositional letters} which can be finite or infinite.


\subsection{Modal logic}

The set ${\agts}$ is a set of indices which can be finite or infinite.
The \emph{multi-modal language} $\langML$  is defined inductively by the following grammar in BNF:
\begin{align*}
\langML:~~\phi\grammardef p\grammarsep \neg p\grammarsep(\phi\land\phi)\grammarsep (\phi\vee\phi)
\grammarsep\Diamond_j\phi\grammarsep\nec_j\phi 
\end{align*}
where $p\in\prop$ and $j\in\agts$. 
\renewcommand{\Kripke}{\mathcal{E}_{\logicML}}
We present the so-called \emph{possible world  semantics} of modal logic.
\label{sec:2.2.2}
A \emph{Kripke model} $\M$ is a tuple $\M\eqdef\left(W,\left\{\R_1,\ldots,\R_m,\ldots,P_1,\ldots,P_n,\ldots\right\}\right)$ where 
\begin{itemize} \index{Possible world} \index{Valuation function} \index{Kripke model ! pointed}
	\item $W$ is a non-empty set whose  elements are called \emph{possible worlds};
	\item $\R_1,\ldots,\R_m,\ldots\subseteq W\times W$ are binary relations over $W$ called \emph{accessibility relations}; 
	\item $P_1,\ldots,P_n,\ldots\subseteq W$ are unary relations  interpreting the propositional letters of $\Prop$.  
\end{itemize}
We  write $w\in \M$ for $w\in W$ by abuse and the  pair $(\M,w)$ is called a \emph{pointed Kripke model}. 
The class of all  pointed Kripke models is denoted $\Kripke$. 



%
%
%
%
%
%
%


\index{Satisfaction relation ! Modal logic}
We define the \emph{satisfaction relation} $\valid\logicML\subseteq\Kripke\times\langML$ inductively by the following \emph{truth conditions}. Below, we write $(M,\w)\models\phi$ for $((M,w),\phi)\in\valid\logicML$. 
For all  $(\M,w)\in\Kripke$, all $\phi,\psi\in\langML$, all $p_i\in \Prop$ and all $j\in\agts$,    
\[\begin{array}{l@{\quad}l@{\quad}l}
(\M,w)\models p_i & \mbox{iff} &  P_i(w) \mbox{ holds};\\
(\M,w)\models \neg p_i & \mbox{iff} &  P_i(w) \mbox{ does not hold};\\
(\M,w)\models(\phi\land\psi) & \mbox{iff} & (\M,w)\models\phi\mbox{ and } (\M,w)\models\psi;\\
(\M,w)\models(\phi\vee\psi) & \mbox{iff} & (\M,w)\models\phi\mbox{ or } (\M,w)\models\psi;\\
(\M,w)\models\Diamond_j\phi & \mbox{iff} & \mbox{there exists } v\in W\mbox{ such that } R_j wv \mbox{ and } (\M,v)\models\phi;\\
(\M,w)\models\nec_j\phi & \mbox{iff} & \mbox{for all } v\in W\mbox{ such that } R_j wv, (\M,v)\models\phi.
\end{array}\]

The triple $\left(\langML,\Kripke,\valid\logicML\right)$ forms a logic, that we call  \emph{modal logic}. Bisimulations for modal logic can be found in \cite{BlaRijVen01}.


%
%
%
%
%
%
%
%

			\renewcommand{\langR}{\mathcal{L}_{\textsf{LC}}}
			\renewcommand{\classup}{\mathcal{E}_{\textsf{LC}}}
			
			\subsection{Lambek calculus}
			
			%
			The \emph{Lambek language $\langR$}  is the  set of formulas  defined inductively  by the following grammar in BNF: 
			\begin{align*}
			\langR :  ~~  \phi \grammardef  
			p  \grammarsep    (\phi\fusions \phi) \grammarsep  (\phi \implybiss \phi) \grammarsep   (\phi\implys \phi)
			\end{align*}			
			where $p\in\AtomProp$.
			A \emph{Lambek model} is a tuple  $\M=\left(\W,\left\{\R,P_1,\ldots,P_n,\ldots\right\}\right)$  where:
			\begin{itemize}
				\item $\W$ is a non-empty set; 
				
				\item $\R\subseteq W\times W\times W$ is a ternary  relation over $\W$;		
				\item  $P_1,\ldots,P_n,\ldots\subseteq W$ are unary relations over $\W$. 
			\end{itemize}
			We write $w\in\M$ for $w\in \W$ by abuse and $(\M,w)$ is called a \emph{pointed Lambek model}. 
			The class of all pointed Lambek models is denoted $\classup$. 
			\renewcommand{\langRa}{\langR}
			We define the \emph{satisfaction relation}   $\valid\logicInt\subseteq\classup\times\langRa$ by the following \emph{truth conditions}. Below, we write $(M,\w)\models\phi$ for $((M,w),\phi)\in\valid\logicLC$. 
			For all   Lambek models  $\M=(\W,\left\{\R,P_1,\ldots,P_n,\ldots\right\})$, all $w\in\M$, all  $\phi,\psi\in\langRa$ and all $p_i\in\Prop$, 
			\[\begin{array}{l@{\quad}l@{\quad}l@{\quad}l@{\quad}l}
			(\M,\xR)\models p_i  & \mbox{ iff } & P_i(x) \mbox{ holds};\\
			%
			(\M,\xR)\models (\phi\fusions \psi) & \mbox{ iff } & \mbox{there are } \yR,\zR\in\W \mbox{ such that } \R\yR\zR\xR,\\
			& & (\M,\yR)\models \phi \mbox{ and } (\M,\zR)\models \psi;\\
			(\M,\xR)\models (\phi\implys \psi) & \mbox{  iff }  & \mbox{for all } \yR,\zR\in\W \mbox{ such that } \R\xR\yR\zR,\\
			& &  \mbox{if } (\M,\yR)\models \phi \mbox{ then } (\M,\zR)\models \psi;\\
			(\M,\xR)\models (\psi\implybiss \phi) & \mbox{ iff } & \mbox{for all } \yR,\zR\in\W \mbox{ such that } \R\yR\xR\zR,\\
			& &  \mbox{if } (\M,\yR)\models \phi\mbox{ then } (\M,\zR)\models \psi. 
			\end{array}\]
			The triple $(\langRa,\classup,\valid\logicLC)$ forms a logic, that we call the \emph{Lambek calculus}. Bisimulations  for the Lambek calculus, called \emph{directed bisimulations}, can be found in \cite{Res00}.

\subsection{Modal intuitionistic logic} 

\label{sec:2.2.3}

 The \emph{modal intuitionistic  language} $\langIL$  is defined inductively by the following grammar in BNF:
\begin{align*}
\langIL:~~\phi\grammardef\top\grammarsep\bot\grammarsep p\grammarsep(\phi\land\phi)\grammarsep (\phi\vee\phi)
\grammarsep(\phi\implyint\phi)\grammarsep\Diamond\phi\grammarsep\nec\phi 
\end{align*}
where $p\in\prop$. A \emph{modal intuitionistic model} is a tuple $M=\left(W,\left\{R,R_{\Diamond},P_1,\ldots,P_n,\ldots\right\}\right)$ where:
\begin{itemize}
	\item $W$ is a non-empty set;
	\item $R\subseteq W\times W$ is a binary relation over $W$ which is reflexive and transitive ($R$ is \emph{reflexive} if  for all $w\in W$ $R w w$ and \emph{transitive} if for all $u,v,w\in W$,  $R u v $ and $R v w$ imply $R u w$);
	\item $R_\Diamond\subseteq W\times W$ is a binary relation over $W$;
	\item  $P_1,\ldots,P_n,\ldots\subseteq W$ are unary relations over $W$ such that for all $v,w\in W$, if $R v w$ and $P_n(v)$ then $P_n(w)$. 
\end{itemize}

We  write $w\in \M$ for $w\in W$ by abuse  and the  pair $(\M,w)$ is called a \emph{pointed modal intuitionistic model}. The class of all pointed modal intuitionistic models is denoted $\classint$. We define the \emph{satisfaction relation}   $\valid\logicInt\subseteq\classint\times\langIL$ by the following \emph{truth conditions}.  Below, we write $(M,\w)\models\phi$ for $((M,w),\phi)\in\valid\logicInt$. 
For all  modal intuitionistic  models  $\M=\left(\W,\left\{\R,R_{\Diamond},P_1,\ldots,P_n,\ldots\right\}\right)$, all $w\in\M$, all  $\phi,\psi\in\langIL$ and all $p_i\in\Prop$, 
\[\begin{array}{l@{\quad}l@{\quad}l@{\quad}l@{\quad}l}
(\M,\xR)\models\top  && \mbox{always};  \\	(\M,\xR)\models\bot &&  \mbox{never};\\
(\M,\xR)\models p_i  & \mbox{ iff } & P_i(w) \mbox{ holds};\\
%
(\M,w)\models(\phi\land\psi) & \mbox{ iff } & (\M,w)\models\phi\mbox{ and } (\M,w)\models\psi;\\
(\M,w)\models(\phi\vee\psi) & \mbox{ iff } & (\M,w)\models\phi\mbox{ or } (\M,w)\models\psi;\\
(\M,\xR)\models (\phi\implyint \psi) & \mbox{ iff } & \mbox{for all } \yR\in\W \mbox{ such that } \R\xR\yR, 
\mbox{if } (\M,\yR)\models \phi \mbox{ then } (\M,\yR)\models \psi;\\
(\M,\xR)\models \Box\phi & \mbox{  iff }  & \mbox{for all } \yR\in\W \mbox{ such that } \R\xR\yR,\\
& &  \mbox{for all  } \zR\in \W \mbox{ such that } \R_{\Diamond} \yR\zR,  (\M,\zR)\models \phi;\\
(\M,\xR)\models \Diamond\phi & \mbox{ iff } & \mbox{for all } \yR\in\W \mbox{ such that } \R\xR\yR,\\
& &  \mbox{there is } \zR\in\W \mbox{ such that } \R_{\Diamond} \yR\zR \mbox{ and } (\M,\zR)\models \phi. 
\end{array}\]

The triple $(\langIL,\classint,\valid\logicInt)$ forms a logic, that we call  \emph{modal intuitionistic logic}. Bisimulations for (modal) intuitionistic logic can be found in \cite{Olk14,Olk17}.

\section{Atomic and molecular logics}

\label{section:4}

\renewcommand{\entailR}{~ {\footnotesize{\dststile{\textsf{}}{}}}~ }


%


\renewcommand{\logicFO}{\textsf{FOL}}

\subsection{Atomic logics}

Atomic logics are  logics such that the  truth conditions of their connectives are  defined by first-order formulas of the   form 
$\forall x_1\ldots x_n(\pm_1 \PP_1 x_1\vee\ldots\vee \pm_n \PP_n x_n\vee \pm \RP x_1\ldots x_n x)$ or $\exists x_1\ldots x_n(\pm_1 \PP_1 x_1 \land\ldots\land\pm_n \PP_n x_n\land \pm \RP x_1\ldots x_n x)$ where $\pm_i$ and $\pm$ are  either empty or $\neg$. Likewise,  propositional letters are defined by first-order formulas of the form $\pm \PP x$. We will represent the structure of these formulas by means of so--called \emph{skeletons} whose various arguments  capture the different features and patterns  that allow us to define  them completely.

We recall that  $\mathbb{N}^*$ denotes the set of natural numbers minus 0 and that for all $n\in\mathbb{N}^*$,  $\mathfrak{S}_{n}$  denotes the group  of permutations over the set $\{1,\ldots,n\}$. Permutations  are generally denoted $\sigma,\tau$,  the identity permutation $\Identity$  is sometimes denoted 1 as the neutral element of every permutation group and $\sigma^-$ stands for the inverse permutation of the permutation $\sigma$. For example, the permutation $\sigma=(3,1,2)$ is the mapping  that maps 1 to 3, 2 to 1 and 3 to 2  (see  for instance \cite{Rot95}  for more details). 


%
%
%
%


\label{sec:IIIB}



\begin{definition}[Atomic  skeletons and connectives]  The  sets of \emph{atomic skeletons}  $\Prop$ and   $\Con$  are defined as follows: 
	\begin{align*}
		\Prop\eqdef& \mathfrak{S}_1\times
		\left\{\direct,\conv\right\}\times\{\forall,\exists\}\times\mathbb{N}^*  \\ 	\Con\eqdef& \Prop\cup\underset{n\in\mathbb{N}^*}{\bigcup}\left\{ \mathfrak{S}_{n+1}\times\left\{\direct,\conv\right\}\times\left\{\forall,\exists\right\}\times{\mathbb{N}^*}^{n+1}\times\left\{+,-\right\}^{n}\right\}. 
	\end{align*} 
	$\Prop$ is called the set of \emph{propositional letter skeletons} and $\Con$ is called the set of \emph{connective skeletons}. 
	\renewcommand{\bb}{\pm}	
	They can  be represented by tuples   $(\sigma,\bb,\fe,\vecmult,\vectoni)$ or $(\sigma,\bb,\fe,\mult)$ if it is a propositional letter skeleton, where    
	$\fe\in\left\{\forall,\exists\right\}$ is called the \emph{quantification signature} of the skeleton,  $\vecmult=(\mult,\mult_1,\ldots,\mult_n)\in{\mathbb{N}^*}^{n+1}$ is called the \emph{type signature} of the skeleton and $\vectoni=(\pm_1,\ldots,\pm_n)\in\left\{+,-\right\}^n$ is called the \emph{tonicity signature} of  the skeleton; $(\fe,\vecmult,\vec{\pm_j})$ is called the \emph{signature} of the skeleton. 
		The \emph{arity} of a propositional letter skeleton  $(\sigma,\bb,\fe,\mult)$ is 0 and its \emph{type} is $\mult$. The \emph{arity} of  a skeleton $\conn\in\Con$   is $n$, its \emph{input types} are $\mult_1,\ldots,\mult_n$ and its \emph{output type} is  $\mult$. 
		%
			
			An \emph{(atomic) connective} or \emph{(atomic) propositional letter} is an object to which is associated an (atomic) skeleton. 
			Its  arity, signature, quantification signature, type signature, tonicity signature, input and output types are the same as its skeleton. By abuse, we sometimes identify a connective with its skeleton. 
			We also introduce the \emph{Boolean connectives} called  \emph{conjunctions and disjunctions}: 
			\begin{align*} \Bool&\eqdef\left\{\land_\mult,\vee_\mult\bmid\mult\in\mathbb{N}^*\right\}
			\end{align*} 
			
			The  type signatures of $\land_\mult$ and $\vee_\mult$ are  $(\mult,\mult,\mult)$ and their arity is 2.
			
			We say that a set of atomic connectives $\Conex$ \emph{is complete for conjunction and disjunction} when  it contains all conjunctions and disjunctions $\land_\mult,\vee_\mult$,  for $\mult$ ranging over all input types and output types of the atomic connectives of $\Conex$.  The set of atomic skeletons associated to $\Conex$ is denoted $\conn(\Conex)$, its set of propositional letters is denoted $\Prop(\Conex)$.

			Propositional letters are denoted $\letterp, p_1, p_2$, \emph{etc}.  and connectives are denoted $\conn, \conn_1, \conn_2$, \emph{etc}.
		\end{definition}


\begin{remark}
	The permutations $\sigma$ mentioned in  atomic skeletons 
	play an important role in the proof theory of atomic logics, which is dealt with in \cite{Auc20,Auc21}.
\end{remark}


\begin{definition}[Atomic language]\label{def:langprotoi}  Let $\Conex$ be a set of atomic connectives. The \emph{(typed) atomic language}  $\langProtoConex$ associated to $\Conex$  is the smallest set that contains the propositional letters  and that is closed under the atomic   connectives. That is, 
	\begin{itemize}
		\item $\Prop(\Conex)\subseteq\langProtoConex$; 
		\item for all  $\conn\in \Conex$ of arity $n>0$ and of type signature $(\mult,\mult_1,\ldots,\mult_n)$  and for all   $\phi_1,\ldots,\phi_n\in\langProtoConex$ of types $\mult_1,\ldots,\mult_n$ respectively,  we have that $\conn(\phi_1,\ldots,\phi_n)\in\langProtoConex$ and $\conn(\phi_1,\ldots,\phi_n)$ is of \emph{type} $\mult$.
	\end{itemize}
	
	The  \emph{Boolean atomic language}  $\langProtoConexP$   is the smallest set that contains the propositional letters  and that is  closed  under the atomic connectives of $\Conex$ as well as  the  Boolean connectives   $\mathbb{B}$:
	\begin{itemize}
		\item for all  $\phi,\psi\in\langProtoConexP$ of type $\mult$, we have that $(\phi\landst_\mult\psi),(\phi\vee_\mult\psi)\in\langProtoConexP$. 
	\end{itemize} 
	Elements of $\langProtoConex$  are 
	 denoted $\phi,\psi,\alpha,\ldots$ The \emph{type of a formula $\phi\in\langProtoConex$} is denoted $\mult(\phi)$.    
	For all $\phi_1,\ldots,\phi_n\in\langProto$ of type $\mult$, $\bigwedge\left\{\phi_1,\ldots,\phi_n\right\}$ and $\bigvee\left\{\phi_1,\ldots, \phi_n\right\}$ stand for $\left(\left(\phi_1\landst_\mult\phi_2\right)\landst_\mult\ldots\landst_\mult\phi_n\right)$ and\linebreak $(\left(\phi_1\veest_\mult\phi_2\right)$ $\veest_\mult\ldots\veest_\mult\phi_n)$ respectively. 
	When it is clear from the context, we will omit the subscript $\mult$ in $\land_\mult,\vee_\mult$ and write them $\land,\vee$.
	
	
	\emph{In the sequel, we assume that all    sets of connectives  $\Conex$ are such that they contain at least a propositional letter}. 
\end{definition}


\begin{definition}[$\Conex$--models]\label{def:C-frame} Let $\Conex$ be a set of atomic connectives. 
	A \emph{$\Conex$--model} is a tuple $\M=(\W,\Rset)$ where 
	$\W$ is a non-empty set  
	and $\Rset$  is a set of  relations over $W$ such that  each $n$--ary connective $\conn\in\Conex$ which is not a Boolean connective  of type signature $(\mult,\mult_1,\ldots,\mult_n)$ is associated to  a $\mult_1+\ldots+\mult_n+\mult$--ary relation $R_{\conn}\in\Rset$. 
	An \emph{assignment} is a tuple $(\xR_1,\ldots,\xR_\mult)\in W^\mult$ for some $\mult\in\mathbb{N}^*$,   generally denoted $\vec{\xR}$. 
	A \emph{pointed $\Conex$--model} $(\M,\vec{\xR})$ is a $\Conex$--model $\M$ together with an assignment $\vec{\xR}$. In that case, we say that $(M,\vec{w})$ \emph{is  of type} $\mult$.  
	The class of all pointed $\Conex$--models is denoted $\classConex$. 
	%
\end{definition}



%
%

\renewcommand{\x}{w}

\begin{definition}[Atomic logics]\label{def:10}  Let  $\Conex$ be a set of atomic connectives and let $\M=(\W,\Rset)$  be a $\Conex$--model.  We define the \emph{interpretation function  of $\langProtoConex$ in  $\M$}, denoted   $\exto{\cdot}^\M:\langProtoConex\rightarrow \underset{\mult\in\mathbb{N}^*}{\bigcup}\W^\mult$, inductively  as follows: for all  propositional letters $\letterp\in\Conex$ of skeleton $(\Identity,\pm,\fe,\mult)$,  all connectives  $\conn\in\Conex$ of skeleton $(\sigma,\pm,\fe,(\mult,\mult_1,\ldots,\mult_n),(\pm_1,\ldots,\pm_n))$ of arity $n>0$ and all $\mult\in\mathbb{N}^*$,  
	for all $\phi,\psi,\phi_1,\ldots,\phi_n\in\langProtoConex$, if $\mult(\phi)=\mult(\psi)=\mult$, 
	\[\begin{array}{r@{\quad}l@{\quad}l@{\quad}l}
	\exto{\letterp}^\M & \eqdef & \bb R_{\letterp}  \\
	\exto{(\phi\landst_\mult\psi)}^\M & \eqdef & \exto{\phi}^\M\cap\exto{\psi}^\M\\
	\exto{(\phi\veest_\mult\psi)}^\M & \eqdef & \exto{\phi}^\M\cup\exto{\psi}^\M\\
	\exto{\conn(\phi_1,\ldots,\phi_n)}^\M  & \eqdef & 
	f_{\conn}(\exto{\phi_1}^\M,\ldots,\exto{\phi_n}^\M) 
	\end{array}\]
	where   $+ R_\letterp\eqdef R_\letterp$ and $-R_\letterp\eqdef W^{\mult}-R_\letterp$ and  
	the function 
	$f_{\conn}$ is defined  
	as follows:
	for all $\W_1\in\mathcal{P}(W^{\mult_1}), \ldots, \W_n\in \mathcal{P}(W^{\mult_n})$, 
	$f_\conn(\W_1,\ldots,\W_n) \eqdef\left\{\vec{\x}_{n+1}\in \W^\mult\bmid \mathcal{C}^{\conn}\left(\W_1,\ldots,\W_n, \vec{\x}_{n+1} \right)  \right\}$
	where 
	$\mathcal{C}^\conn\left(\W_1,\ldots,\W_n, \vec{\x}_{n+1}\right)$ is called the \emph{truth condition} of $\conn$ and is defined as follows:
	\begin{itemize}
		\item 
		if $\fe=\forall$: 	  
		``$\forall\vec{\x}_1\in\W^{\mult_1}\ldots\vec{\x}_n\in\W^{\mult_n} \left( \vec{\x}_1\pitchfork_1\W_1\vee\ldots\vee\vec{\x}_n\pitchfork_n\W_n\vee R_\conn^{\pm\sigma}\vec{\x}_1\ldots\vec{\x}_n\vec{\x}_{n+1}\right)$'';

		\item  if $\fe=\exists$: 
		``$\exists\vec{\x}_1\in\W^{\mult_1}\ldots\vec{\x}_n\in \W^{\mult_n}\left( \vec{\x}_1\pitchfork_1\W_1\land\ldots\land\vec{\x}_n\pitchfork_n\W_n\land R_\conn^{\pm\sigma}\vec{\x}_1\ldots\vec{\x}_n\vec{\x}_{n+1}\right)$'';
	\end{itemize}
	where, for all  $j\in\exto{1;n}$, 				
	$\vec{\x}_j\pitchfork_j\W_j\eqdef\begin{cases}
	\vec{\x}_j\in \W_j & \mbox{if $\pm_j=+$}\\  
	\vec{\x}_j\notin \W_j & \mbox{if $\pm_j=-$}\end{cases}$ and $R_\conn^{\pm\sigma}\vec{\x}_1\ldots\vec{\x}_{n+1}$ holds iff \linebreak$\pm R_\conn\vec{\x}_{\sigma^-(1)}\ldots\vec{\x}_{\sigma^-(n+1)}$ with the notations  $+ R_\conn\eqdef R_\conn$ and $-R_\conn\eqdef W^{\mult+\mult_1+\ldots+\mult_n}-R_\conn$. 	
	%
	If  $\mathcal{E}_\Conex$ is a class of pointed  $\Conex$--models, 
	the \emph{satisfaction relation} $\entailR\subseteq \mathcal{E}_\Conex\times\langProtoConex$ is defined as follows:  for all $\phi\in\langProtoConex$ and all $(\M,\vec{\xR})\in\mathcal{E}_\Conex$,   $\left((\M,\vec{\xR}),\phi\right)\in\entailR$ iff  $\vec{\xR}\in \exto{\phi}^\M$. We usually write $(\M,\vec{\xR})\entailR\phi$ instead of  $\left((\M,\vec{\xR}),\phi\right)\in\entailR$ and we say that $\phi$ is \emph{true} in $(M,\vec{\xR})$. 
	
	The class of \emph{atomic logics} is defined by $\classGaggleLogics\eqdef\{(\langProtoConex,\classConexE,\entailR)\bmid \Conex\mbox{ is a finite set of  atomic connectives}$ $\mbox{and } \classConexE \mbox{ is a class of $\Conex$--models}\}$. 
	The atomic logic $(\langProtoConex,\classConexE,\entailR)$ is  the \emph{atomic logic associated to  $\classConexE$ and $\Conex$}. The logics of the form   $(\langProtoConex,\classConex,\entailR)$ are    called  \emph{basic atomic logics}. We call them \emph{Boolean (basic) atomic logics} when their language includes the Boolean connectives $\Bool$. 
\end{definition}


%

\renewcommand{\x}{x}

\begin{example}[Lambek calculus, modal logic] \label{example:6}
	The Lambek calculus, 
	where $\Conex=\{p,\circ,\backslash,\slash\}$ is defined in Section \ref{sec:2.2.1}, is an example of   atomic logic. Here $\circ,\backslash,\slash$ are the connectives  of skeletons $(\sigma_1,+,\signature_1),$ $(\sigma_5,-,\signature_3),(\sigma_3,-,\signature_2)$. Another example of atomic logic is  modal logic 
	where  $\Conex=\{p,\top,\bot,\land,\vee,\Diamond,\Box\}$ is   such that
	\begin{itemize}
		\item $\top,\bot$ are  connectives of skeletons $(\Identity,+,\exists,1)$ and $(\Identity,-,\forall,1)$ respectively;  
		\item $\land,\vee,\Diamond,\Box$ are  connectives of skeletons     $(\sigma_1,+,\signature_1)$, $(\sigma_1,-,\signature_4)$, $(\tau_2,+,\signatureb_1)$ and  $(\tau_2,-,\signatureb_2)$ respectively; 
		\item   $\Conex$-models $\M=(\W,\Rset)\in\mathcal{E}_{\Conex}$  are such that $R_{\land}=R_{\vee}=\{(\w,\w,\w)\bmid \w \in\W\}$, $R_{\Diamond}=R_{\Box}$ and  $R_{\top}=R_{\bot}=\W$. 
	\end{itemize}
	Indeed, one can easily show that, with these conditions on the $\Conex$--models of  $\mathcal{E}_{\Conex}$, we have that for all $(\M,w)\in\mathcal{E}_{\Conex}$,  $(\M,\w)\entailR \land(\phi,\psi)$ iff $(\M,\w)\entailR \phi$ and $(\M,\w)\entailR\psi$, and $(\M,\w)\entailR \vee(\phi,\psi)$ iff $(\M,\w)\entailR \phi$  or $(\M,\w)\entailR \psi$. 
	The Boolean conjunction and disjunction $\landst$ and $\veest$ are defined using the connectives of $\Con$ by means of special relations $R_\landst$ and $R_\veest$. However, they could obviously be defined directly. 
	Many more examples of atomic connectives  are   in Figure  \ref{figure:1}. They are in fact just examples of gaggle connectives since all gaggle logics \cite{Auc20,Auc21}  are also atomic logics; they are all of type signature $(1,1,\ldots,1)$.  All the possible truth conditions of  unary and binary  atomic connectives of this type signature are   in \cite{Auc20,Auc21}. 
\end{example}

\begin{figure}
	\begin{center}
\parbox{8cm}{\begin{tabular}{|l||l|}
			\toprule
			Permutations of $\mathfrak{S}_2$ & unary signatures \\ 
			\midrule
			$\tau_1$ $=(1,2)$ & $\signatureb_1=(\exists,(1,1),+)$\\
			$\tau_2$  $=(2,1)$ & $\signatureb_2=(\forall,(1,1),+)$\\
			\cline{2-2}
			& $\signatureb_3=(\forall,(1,1),-)$\\
			\cline{2-2}
			& $\signatureb_4=(\exists,(1,1),-)$\\
			\bottomrule
				\end{tabular}}
\parbox{8cm}{			
				\begin{tabular}{|l||l|}
					\toprule
			Permutations of $\mathfrak{S}_3$ & binary signatures \\ 
			\midrule
			$\sigma_1$ $=(1,2,3)$ & $\signature_1=\left(\exists,(1,1,1),\left(+,+\right)\right)$\\
			$\sigma_2$  $=(3,2,1)$ & $\signature_2=\left(\forall,(1,1,1),\left(+,-\right)\right)$\\
			$\sigma_3$  $=(3,1,2)$ & $\signature_3=\left(\forall,(1,1,1),\left(-,+\right)\right)$\\
			\cline{2-2}
			$\sigma_4$  $=(2,1,3)$ & $\signature_4=\left(\forall,(1,1,1),\left(+,+\right)\right)$\\
			$\sigma_5$  $=(2,3,1)$ & $\signature_5=\left(\exists,(1,1,1),\left(+,-\right)\right)$\\
			$\sigma_6$  $=(1,3,2)$ & $\signature_6=\left(\exists,(1,1,1),\left(-,+\right)\right)$\\
			\cline{2-2}
			& $\signature_7=\left(\exists,(1,1,1),\left(-,-\right)\right)$\\
			\cline{2-2}
			& $\signature_8=\left(\forall,(1,1,1),\left(-,-\right)\right)$\\
			\bottomrule
		\end{tabular}}\caption{Permutations of $\mathfrak{S}_2$ and $\mathfrak{S}_3$ and `orbits' of unary and  binary signatures \label{fig:unary conn}}	
	\end{center}
\end{figure}

\renewcommand{\x}{w} 
\renewcommand{\y}{v}
\renewcommand{\z}{u}

\begin{figure}
	\begin{center}
		\begin{tabular}{|l|l|l|}
			\toprule
			Atomic connective  & Truth condition & Non--classical  con. \\
			&  & in the literature \\
			\midrule
			\multicolumn{3}{c}{The conjunction orbit}\\
			\midrule
			$\phi\cmmm\psi$  & $\exists \y \z\left(\y\in\exto{\phi}\land \z\in\exto{\psi}\land R\y\z\x \right)$ & $\phi\circ\psi$  \cite{Lam58}, $\phi\fusion\psi$ \cite{Auc14c} \\ 
			$\phi\cmmmpmp\psi$  & $\forall \y\z\left(\y\in\exto{\phi}\vee\z\notin\exto{\psi}\vee\complem{R}\x\z\y \right)$ & \\ 
			$\phi\cmmmpmpb\psi$ & $\forall\y\z \left(\y\in\exto{\phi}\vee\z\notin\exto{\psi}\vee\complem{R}\z\x\y \right)$ & $\slash$ \cite{Lam58},  $\phi\implybis\psi$ \cite{Auc14c} \\
			$\phi\cmmmb\psi$ & $\exists\y\z\left(\y\in\exto{\phi}\land \z\in\exto{\psi}\land R\z\y\x \right)$ & \\ 
			$=\psi\cmmm\phi$ & & \\
			$\phi\cmmmmpp\psi$ & $\forall\y\z\left(\y\notin\exto{\phi}\vee\z\in\exto{\psi}\vee\complem{R}\x\y\z \right)$ &  $\backslash$ \cite{Lam58},  $\phi\implyun\psi$ \cite{Auc14c}\\
			$=\psi\cmmmpmp\phi$ & & \\
			$\phi\cmmmmppb\psi$ & $\forall\y\z\left(\y\notin\exto{\phi}\vee\z\in\exto{\psi}\vee\complem{R}\y\x\z \right)$ & \\ 
			$=\psi\cmmmpmpb\phi$ & & \\
			\midrule
			\multicolumn{3}{c}{The not--but  orbit}\\
			\midrule
			$\phi\cpmm\psi$  & $\exists \y \z\left(\y\notin\exto{\phi}\land \z\in\exto{\psi}\land R\y\z\x \right)$ & $\phi\RAtrois\psi$ \cite{Auc14c}\\
			$\phi\cpmmpmm\psi$ & $\exists \y \z\left(\y\notin\exto{\phi}\land \z\in\exto{\psi}\land R\x\z\y \right)$ & \\ 
			$\phi\cpmmppp\psi$  & $\forall \y\z\left(\y\in\exto{\phi}\vee\z\in\exto{\psi}\vee\complem{R}\z\x\y \right)$ & $\phi\fissiondeux\psi$  \cite{Auc14c}\\ 
			$\phi\cpmmmpm\psi$  & $\exists \y\z\left(\y\in\exto{\phi}\land\z\notin\exto{\psi}\land R\z\y\x \right)$ & \\ 
			$=\psi\cpmm\phi$ & & \\
			$\phi\cpmmmpmb\psi$  & $\exists \y\z\left(\y\in\exto{\phi}\land\z\notin\exto{\psi}\land R\x\y\z \right)$ & $\phi\BRAun\psi$ \cite{Auc14c}\\ 
			$=\psi\cpmmpmm\phi$ & & \\
			$\phi\cpmmpppb\psi$  & $\forall \y\z\left(\y\in\exto{\phi}\vee\z\in\exto{\psi}\vee \complem{R}\y\x\z \right)$ & \\ 
			$=\psi\cpmmppp\phi$ & & \\
			\midrule
			\multicolumn{3}{c}{The but--not  orbit}\\
			\midrule
			$\phi \cmpm\psi$  & $\exists \y \z\left(\y\in\exto{\phi}\land \z\notin\exto{\psi}\land R\y\z\x \right)$ & $\phi\BRAtrois\psi$  \cite{Auc14c}\\ 
			$\phi \cmpmppp\psi$ & $\forall \y \z\left(\y\in\exto{\phi}\vee \z\in\exto{\psi}\vee \complem{R}\x\z\y \right)$ & \\ 
			$\phi \cmpmpmm\psi$  & $\exists \y\z\left(\y\notin\exto{\phi}\land\z\in\exto{\psi}\land R\z\x\y \right)$ & $\phi\RAdeux\psi$  \cite{Auc14c} \\ 
			$\phi \cmpmpmmb\psi$  & $\exists \y\z\left(\y\notin\exto{\phi}\land\z\in\exto{\psi}\land R\z\y\x \right)$ &  $\phi\varobslash\psi$ \cite{Gri83,Moo07} \\ 
			$=\psi \cmpm\phi$ & & \\
			$\phi \cmpmpppb\psi$  & $\forall \y\z\left(\y\in\exto{\phi}\vee\z\in\exto{\psi}\vee \complem{R}\x\y\z \right)$ & $\phi\varoplus\psi$ \cite{Gri83,Moo07}\\ $=\psi \cmpmppp \phi$ & &  $\phi\fissionun\psi$  \cite{Auc14c} \\ 
			$\phi \cmpmmpmb \psi$  & $\exists \y\z\left(\y\in\exto{\phi}\land \z\notin\exto{\psi}\land R\y\x\z \right)$ & $\phi\varoslash\psi$ \cite{Gri83,Moo07} \\ 
			$=\psi \cmpmpmm\phi$ & &\\
			\midrule
			\multicolumn{3}{c}{The stroke  orbit}\\
			\midrule
			$\phi \cppm \psi$  & $\exists  \y \z\left(\y\notin\exto{\phi}\land \z\notin\exto{\psi}\land R\y\z\x \right)$ & $\phi\NANDtrois\psi$ \cite{AllDun93,Gor98} \\ 
			$\phi \cppmppm \psi$ & $\exists \y \z\left(\y\notin\exto{\phi}\land \z\notin\exto{\psi}\land R\x\z\y \right)$ & \\ 
			$\phi \cppmppmb \psi$  & $\exists \y\z\left(\y\notin\exto{\phi}\land\z\notin\exto{\psi}\land R\z\x\y \right)$ & \\ 
			$\phi \cppmppmc \psi$  & $\exists \y\z\left(\y\notin\exto{\phi}\land\z\notin\exto{\psi}\land R\z\y\x \right)$ & \\ 
			$=\psi \cppm \phi$ & &\\
			$\phi \cppmppmd\psi$  & $\exists \y\z\left(\y\notin\exto{\phi}\land\z\notin\exto{\psi}\land R\x\y\z \right)$ &  $\phi\NANDun\psi$ \cite{AllDun93,Gor98} \\ 
			$=\psi \cppmppm\phi$ & &\\
			$\phi \cppmppme \psi$  & $\exists \y\z\left(\y\notin\exto{\phi}\land\z\notin\exto{\psi}\land R\y\x\z \right)$ & $\phi\NANDdeux\psi$ \cite{AllDun93,Gor98}\\ 
			$=\psi \cppmppmb \phi$ & &\\
			\bottomrule
		\end{tabular}\caption{Some binary  connectives of  atomic logics of type $(1,1,1)$  \label{figure:1}}
	\end{center}
\end{figure}

\renewcommand{\x}{w} 
\renewcommand{\y}{v}
\renewcommand{\z}{u}

\subsection{Molecular logics}

Molecular logics are  logics whose primitive connectives are compositions of atomic connectives. That is why we call them `molecular', just as molecules are compositions of atoms in chemistry.

\begin{definition}[Molecular skeleton and  connective] The class $\Con^*$ of \emph{molecular skeletons} is the smallest set such that:
	
	\begin{itemize}
		\item $\Prop\cup\Bool\subseteq\Con^*$  and $\Con^*$ contains for each $\mult\in\mathbb{N}^*$ a symbol $\id_{\mult}$ of \emph{type signature} $(\mult,\mult)$ and \emph{arity} 1; 
		\item for all $\conn\in\Con$ of type signature $(\mult,\mult_1,\ldots,\mult_n)$  and all $\comc_1,\ldots,\comc_n\in \Con^*$ of  type signatures   $(\mult_1,\mult_1^1,\ldots,\mult_{a_1}^1),$ $\ldots,(\mult_n,\mult_1^n,\ldots,\mult_{a_n}^n)$ respectively, the connective $\conn(\comc_1,\ldots,\comc_n)$ belongs to $\Con^*$,  its \emph{type signature} is $(\mult,\mult_1^1,\ldots,\mult^1_{a_1},\ldots,\mult_1^n,\ldots,\mult_{a_n}^n)$ and its  \emph{arity}  is $a_1+\ldots+a_n$.  
	\end{itemize}
	We  define the \emph{quantification signature} $\fe(\comc)$ of $\comc=\conn(\comc_1,\ldots,\comc_n)$ by $\fe(\comc)\eqdef\fe(\conn)$.
	
	If $\comc\in\Con^*$, we define its \emph{decomposition tree} as follows. If $\comc=\conn\in\Con$ is of arity $n>0$, then its decomposition tree $T_\comc$ is the tree of root $\conn$  with $n$ children--leaves   labeled by $\id$.
	If $\comc=\conn(\comc_1,\ldots,\comc_n)\in\Con^*$ then  its decomposition tree $T_\comc$ is a  tree labeled with atomic connectives defined inductively as follows: the  root of $T_\comc$ is $\comc$ and it is labeled  with $\conn$ and one sets edges between that  root and the roots $\comc_1,\ldots,\comc_n$ of the decomposition trees $T_{\comc_1},\ldots, T_{\comc_n}$ respectively.   

	A \emph{molecular connective}  is an object to which is associated a molecular skeleton. Its arity, quantification signature and decomposition tree are the same as its skeleton.

	The set of \emph{atomic connectives associated   to a set $\Conex$ of molecular connectives} is the set of labels  different from $\id$ of the decomposition trees of the molecular connectives of $\Conex$.
\end{definition}



\renewcommand{\x}{x}
\renewcommand{\y}{y}
\renewcommand{\z}{z}

\begin{example}[Modal intuitionistic logic]\label{example:9}
	Let us consider the connectives defined by the following first--order formulas:  
	\begin{align*}
	\comc(\x)\eqdef\forall \y&\left(\RP\x\y\rightarrow\forall \z \left(\RP_\Diamond \y\z\rightarrow  \PP(\z) \right)\right)\\
	\comc'(\x)\eqdef\forall \y&\left(\RP\x\y\rightarrow\exists\z\left(\RP_{\Diamond}\y\z\land \PP(\z) \right)\right)\\
	\conn_1(\x)\eqdef\forall \y & \left(\RP\x\y\rightarrow \PP(\y) \right)\\ 
	\conn_2(\x)\eqdef \forall \z&\left(\RP_\Diamond\y\z\rightarrow \PP(\z) \right) \\ 
	\conn_3(\x)\eqdef\exists\z&\left(\RP_{\Diamond}\y\z\land \PP(\z)\right)   
	\end{align*}
	Then, $\conn_1,\conn_2,\conn_3$ are atomic connectives and the connectives associated to  $\comc,\comc'$ are molecular connectives.  Indeed, $\comc$ is  the composition of $\conn_1$  and $\conn_2$,  $\comc=\conn_1(\conn_2)$,  and $\comc'$ is the composition of $\conn_1$ and $\conn_3$,   $\comc'=\conn_1(\conn_3)$. Equivalently, $\comc$ and $\comc'$ will have the same semantics as  $\comc=\conn_1(\conn_2(\id_1))$ and $\comc'=\conn_1(\conn_3(\id_1))$.  The connective associated to  $\comc$ corresponds to the connective $\Box$ of modal intuitionistic logic and the connective associated to  $\comc'$ corresponds to the connective $\Diamond$ of modal intuitionistic logic \cite{Olk17} defined in Section \ref{sec:2.2.1}. 
\end{example}

\begin{definition}[Molecular language] Let $\Conex$ be a set of molecular connectives. The \emph{(typed) molecular language}  $\langProtoConex$ associated to $\Conex$  is the smallest set that contains the propositional letters  and that is closed under the molecular   connectives while respecting the type constraints. That is, 
	\begin{itemize}
		\item the propositional letters  of $\Conex$ belong to $\langProtoConex$; 
		\item for all  $\conn\in \Conex$  of type signature $(\mult,\mult_1,\ldots,\mult_n)$  and for all   $\phi_1,\ldots,\phi_n\in\langProtoConex$ of types $\mult_1,\ldots,\mult_n$ respectively,  we have that $\conn(\phi_1,\ldots,\phi_n)\in\langProtoConex$ and $\conn(\phi_1,\ldots,\phi_n)$ is of \emph{type} $\mult$.
	\end{itemize}
	
	The  \emph{Boolean molecular language}  $\langProtoConexP$   is the smallest set that contains the propositional letters  and that is  closed  under the molecular connectives of $\Conex$ as well as  the  Boolean connectives   $\mathbb{B}$:
	\begin{itemize}
		\item for all  $\phi,\psi\in\langProtoConexP$ of type $\mult$, we have that $(\phi\landst_\mult\psi),(\phi\vee_\mult\psi)\in\langProtoConexP$. 
	\end{itemize} 

	We say that $\Conex$ \emph{is complete for conjunction and disjunction} when its associated set of atomic connectives	is complete for conjunction and disjunction.

	Elements of $\langProtoConex$  are called \emph{molecular formulas} and  are  denoted $\phi,\psi,\alpha,\ldots$ The \emph{type of a formula $\phi\in\langProtoConex$} is denoted $\mult(\phi)$.    
	We use the same abbreviations as for the atomic language. 
\end{definition}

\begin{definition}[Molecular logic] 	
	If $\Conex$ is a set of molecular connectives, then a \emph{$\Conex$--model} $\M$  is a $\Conex'$--model $\M$ where $\Conex'$ is the set of atomic connectives associated to  $\Conex$. We also define $\Assign\eqdef\vec{w}(\M,\Conex')$.   The truth conditions for molecular connectives are defined naturally inductively  from the truth conditions of atomic connectives of Definition \ref{def:10} (we only give the new cases): for all $\mult\in\mathbb{N}^*$, 
	\[\begin{array}{r@{\quad}l@{\quad}l@{\quad}l}
		\exto{\id_{\mult}(\phi)}^\M&\eqdef&\exto{\phi}^\M\\
		\exto{\conn(\comc_1,\ldots,\comc_n)\left(\phi^1_1,\ldots,\phi_1^{k_1},\ldots,\phi^1_n,\ldots,\phi^{k_n}_n\right)}^\M &  \eqdef &
		f_{\conn}\left(\exto{\comc_1(\phi_1^1,\ldots,\phi^{k_1}_1)}^\M,\ldots,\exto{\comc_n(\phi^1_n,\ldots,\phi^{k_n}_n}^\M\right) 
	\end{array}\]
	If $\classConexE$ is a class of pointed $\Conex$--models, the triple $(\langProtoConex,\classConexE,\entailR)$ is a  logic called the \emph{molecular logic associated to  $\classConexE$ and $\Conex$}. The logics of the form   $(\langProtoConex,\classConex,\entailR)$ are    called  \emph{basic molecular logics}. We call them \emph{Boolean (basic) molecular logics} when their language includes the Boolean connectives $\Bool$. 
\end{definition}


\subsection{Boolean negation}

\label{sec:4.2.3} 

\renewcommand{\x}{x}
\renewcommand{\y}{y}
\renewcommand{\z}{z}

Note that atomic logics do not include Boolean negation as a primitive connective. It turns out that Boolean negation can be defined systematically for each atomic connective by applying a transformation on it. The Boolean negation of a formula then boils down to taking the Boolean negation of the  outermost connective of the formula. This transformation is defined as follows.

\begin{definition}[Boolean negation]\label{def:15} Let $\conn$ be a $n$--ary connective of skeleton$(\sigma,\bb,\fe,\vecmult,\pm_1,\ldots,\pm_n)$. The \emph{Boolean negation of $\conn$} is the connective $-\conn$ of skeleton $(\sigma,-\bb,-\fe,\vecmult,-\pm_1,\ldots,-\pm_n)$ where $-\fe\eqdef\exists$ if $\fe=\forall$ and $-\fe\eqdef\forall$ otherwise, which is associated in any $\Conex$--model to the same relation as $\conn$. If $\phi=\conn(\phi_1,\ldots,\phi_n)$ is an atomic  formula, the \emph{Boolean negation of $\phi$} is the formula $-\phi\eqdef-\conn(\phi_1,\ldots,\phi_n)$.  
\end{definition}

%
%
%
%
\begin{proposition}\label{prop:5b}
	Let $\Conex$ be a set of atomic connectives  such that $-\conn\in\Conex$ for all $\conn\in\Conex$. 
	Let  $\phi\in\langProtoConex$ and  $\M$ be a $\Conex$--model. Then, for all $\vec{w}\in\Assign$, 
	$	\vec{w}\in\exto{-\phi}^\M$ iff $\vec{w}\notin\exto{\phi}^\M$.
\end{proposition}

\section{Automatic  bisimulations for atomic and molecular  logics}

\label{sec:6}

In this section, we are going to see that notions of bisimulations can be automatically defined for atomic logics on the basis of the definition of the truth conditions of their connectives, not only for plain  atomic logics but also for molecular logics. 

\subsection{Atomic logics}

\renewcommand{\x}{w}
\renewcommand{\y}{v}
\renewcommand{\z}{u}

\begin{definition}[$\Conex$--bisimulation]\label{def:inv}
	Let $\Conex$ be a set of atomic connectives,  
	let  $\conn\in\Conex$ and let $M_1=(W_1,\Rset_1)$ and $M_2=(W_2,\Rset_2)$ be two $\Conex$--models. A binary relation $\simproto\subseteq\underset{\mult\in\mathbb{N}^*}{\bigcup}(W_1^\mult\times W_2^{\mult})\cup(W_2^{\mult}\times W_1^\mult)$  is a  \emph{$\Conex$--bisimulation} between $M_1$ and $M_2$  when for all $\conn\in\Conex$, if $\{M,M'\}=\{M_1,M_2\}$, then for  all $\vec{\x}_1,\ldots, \vec{\x}_n,\vec{\x'}_1,\ldots, \vec{\x'}_n,\vec{\x},\vec{\x'}\in \Assign\cup\Assignb$,  
	\begin{enumerate}
		\item if $\conn$ is an propositional letter $p$ then, if  $\vec{\x}\simproto \vec{\x'}$ and  $\vec{\x}\in\exto{p}$  then $\vec{\x'}\in\exto{p}$;
		\item if $\conn$ has skeleton $(\sigma,\pm,\exists, \vecmult,(\pm_1,\ldots,\pm_n))$ and we have   
		$\vec{\x}\simproto\vec{\x'}$	and $R^{\bb\sigma}_{\conn}\vec{\x}_1\ldots\vec{\x}_n\vec{\x}$, then 
		
		$\exists\vec{\x'}_1,\ldots, \vec{\x'}_n\left(		
		\vec{\x}_1\pitchforkb\vec{\x'}_1\land\vec{\x}_2\pitchforkb\vec{\x'}_2\land\ldots\land\vec{\x}_n\pitchforkb\vec{\x'}_n\land R^{'\bb\sigma}_{\conn}\vec{\x'}_1\ldots\vec{\x'}_n\vec{\x'}\right)$;
		\item if $\conn$ has skeleton $(\sigma,\pm,\forall, \vecmult,(\pm_1,\ldots,\pm_n))$ and we have   
		$\vec{\x}\simproto\vec{\x'}$ and  $-R^{'\bb\sigma}_{\conn}\vec{\x'}_1\ldots\vec{\x'}_n\vec{\x'}$, then 
		
		$\exists\vec{\x}_1,\ldots,\vec{\x}_n\left(\vec{\x}_1\pitchforkb\vec{\x'}_1\land \vec{\x}_2\pitchforkb\vec{\x'}_2\land\ldots\land\vec{\x}_n\pitchforkb\vec{\x'}_n\land -R^{\bb\sigma}_{\conn}\vec{\x}_1\ldots\vec{\x}_n\vec{\x}\right)$;  		
	\end{enumerate} 
	where,  for all   $j\in\exto{1;n}$, 	we define 			
	$\vec{\x}_j\pitchforkb\vec{\x'}_j\eqdef \begin{cases}
	\vec{\x}_j\simproto\vec{\x'}_j & \mbox{if $\pm_j=+$}\\
	\vec{\x'}_j\simproto \vec{\x}_j & \mbox{if $\pm_j=-$} 
	\end{cases}$. 
	
	
	When such a $\Conex$--bisimulation  $\simproto$ exists and  $\vec{\x}\simproto\vec{\x'}$, we say that $(M,\vec{\x})$ and $(M',\vec{\x'})$ are \emph{$\Conex$--bisimilar} and we write it  $(M,\vec{\x})\simul_{\Conex}(M',\vec{\x'})$. 
\end{definition}

Note that  case 1. is a  particular instance of cases 2. and 3. with $n=0$. The fact that the order $M_1-M_2$ can be  possibly reversed at the level of the definition is reminiscent of the way \emph{directed} bisimulations are defined for the Lambek calculus, as we will see in  Example \ref{ex:10}. 

\begin{definition}
	Let $\Conex$ be a set of atomic connectives. 
	Let $(M,\vec{\x})$ and $(M',\vec{\x'})$ be two pointed $\Conex$--models. We write $(M,\vec{\x})\equivLang_{\Conex}(M',\vec{\x'})$ when for all $\phi\in\langProtoConex$, $(M,\vec{\x})\entailR\phi$ implies  $(M',\vec{\x'})\entailR\phi$. 
\end{definition}

\begin{proposition}\label{prop:7}
	Let $\Conex$ be a set of atomic connectives  and  let $M_1=(W_1,\Rset_1)$ and $M_2=(W_2,\Rset_2)$ be two $\Conex$--models. Let $\simproto$ 
	be a $\Conex$--bisimulation between $M_1$ and $M_2$. 
	Then, if  $\{M,M'\}=\{M_1,M_2\}$ 
	then for all   $\vec{\x}\in\Assign$, all $\vec{\x'}\in \Assignb$,
	if $\vec{\x}\simproto\vec{\x'}$ then  
	$(\M,\vec{\x})\equivLang_{\Conex}(\M',\vec{\x'})$. 
\end{proposition}

\begin{example}[Modal logic]\label{ex:9}
	Let us consider the connectives of modal logic:  $\Conex=\left\{p,\neg p,\land,\vee,\Diamond,\Box\right\}$ where $p$ has skeleton $(\Identity,+,\exists,1)$, $\neg p$ has skeleton $(\Identity,-,\forall,1)$, $\Diamond$ has skeleton $\cppmmr$ and $\Box$ has skeleton $(\tau_2,-,t_2)$. Let $M_1=(W_1,\left\{R_1,P_1\right\})$ and $M_2=(W_2,\left\{R_2,P_2\right\})$ be two Kripke models (they are also $\Conex$-models). 
	A binary relation  $\simproto$ between $M_1$ and $M_2$ is a  $\Conex$--bisimulation between  $M_1$  and $M_2$  when for all $\M,\M'\in\{\M_1,\M_2\}$ with $M=(W,\{R,P\})$ and $M'=(W',\{R',P'\})$, all  $\x,\y\in\M$ and all $\x',\y'\in\M'$, 
	\begin{itemize}
		\item  
		if $\x\simproto \x'$ and $\x\in \exto{p}$  then $\x'\in\exto{p}$ (condition for $p$); 
		\item   
		if $\x\simproto \x'$ and $\x'\in \exto{p}$  then $\x\in\exto{p}$ (condition for $\neg p$);
		\item    
		if $\x\simproto\x'$ and  $R \x \y$  then there is $ \y'\in W'$ such that $\y\simproto\y'$ and $R'\x'\y' $ (condition for $\Diamond=\cppmmr$);
		\item   
		if $\x\simproto\x'$ and $R'\x'\y'$  then there is $\y\in W$ such that  $\y\simproto\y'$ and $\R\x\y$ (condition for $\Box=\cmmpp$). 
	\end{itemize}
	Note that every  $\Conex$--bisimulation can be canonically extended into a \emph{symmetric} $\Conex$--bisimulation: one sets $\x'\simproto \x$ when $\x\simproto \x'$ already holds.    
\end{example} 

\begin{example}[Lambek calculus]\label{ex:10}
	Let us consider the connectives of the Lambek calculus: $\Conex=\left\{p,\circ,\backslash, \slash\right\}$ where $p$ has skeleton $(\Identity,+,\exists,1)$, $\circ$ has skeleton $\cmmm$, $\backslash$ has skeleton $\cmmmmpp$ and $\slash$ has skeleton $(\sigma_3,-,s_2)$. Let $M_1=(W_1,\{R_1,P_1\})$ and $M_2=(W_2,\{R_2,P_2\})$ be two Lambek models (they are also  $\Conex$--models). 
	A binary relation  $\simproto$ between $M_1$ and $M_2$ is a  $\Conex$--bisimulation between $M_1$ and $M_2$  when for all $\M,\M'\in\{\M_1,\M_2\}$ with $M=(W,\{R,P\})$ and $M'=(W',\{R',P'\})$,   all $\x,\y,\z\in\M$ and all $\x',\y',\z'\in\M'$, 
	\begin{itemize}
		\item  
		if $\x\simproto \x'$ and  $\x\in \exto{p}$  then  $\x'\in\exto{p}$ (condition for $p$); 
		\item  
		if $\x\simproto\x'$ and $R \y\z\x$  then there are $\y',\z'\in W'$ such that $\y\simproto\y'$, $\z\simproto\z'$ and $R\y'\z'\x'$ (condition for $\circ=(\sigma_1,+,s_1)$);
		\item  
		if $\y\simproto\y'$ and $\R\y'\z'\x'$  then there are $\z,\x\in W$ such that $\z'\simproto\z$, $\x\simproto\x'$ and $R\y\z\x$ (condition for $\backslash=$ $\cmmmmpp$);
		\item  
		if $\z\simproto\z'$ and $\R\y'\z'\x'$  then there are $\y,\x\in W$ such that $\y'\simproto\y$,  $\x\simproto\x'$ and  $\R\y\z\x$ (condition for $\slash=\cmmmpmpb$). \qedhere
	\end{itemize}
\end{example}

The following proposition shows that the notions of  $\Conex$--bisimulation for the Lambek calculus and directed bisimulation coincide (directed bisimulations are defined for example  in \cite[Definition~13.2]{Res00}) and likewise for modal logic. 

\begin{proposition}\label{prop:8}
	\begin{itemize}
		\item Let $\Conex=\left\{p,\neg p,\land,\vee,\Diamond,\Box\right\}$ be the connectives  of Example \ref{ex:9} and let $\M$ and $\M'$ be two $\Conex$--models. Then, a $\Conex$--bisimulation  between $\M$ and $\M'$ is a modal bisimulation between $\M$ and $\M'$ and vice versa. 
		\item 
		Let 	$\Conex=\left\{p,\circ,\backslash, \slash\right\}$  be the connectives of Example \ref{ex:10} and let $M$ and $M'$ be two $\Conex$--models. 
		Then, 
		a  $\Conex$--bisimulation between  $\M$ and $\M'$ 
		is a directed bisimulation between $M$ and $M'$ and vice versa. 
	\end{itemize}
\end{proposition}


\begin{example}[Intuitionistic logic]\label{example:7} Let us consider the connectives of intuitionistic 
	logic: 
	$\Conex=\{\letterp,\bot,\top,\land,\vee,$ $\implyint\}$ where $\letterp$ has skeleton $(\Identity,+,\exists,1)$, $\top$ has skeleton $(\Identity,+,\exists,1)$, $\bot$ has skeleton $(\Identity,-,\forall,1)$, $\land$ and $\vee$ are Boolean connectives and $\implyint$ has skeleton $(\sigma_5,-,\signature_3)$ (here, $\top$ and $\bot$ are represented 
	by specific propositional letters of respective signatures $(\Identity,+,\exists,1)$ and $(\Identity,-,\forall,1)$). Let 
	$\M_1=(W_1,\R_1,P)$ and $\M_2=(W_2,\R_2,P)$ be two intuitionistic models. Following the 
	results of \cite[Section~8]{Auc14c}, we represent these intuitionistic models by the 
	$\Conex$--models $\M_1^{\implyint}=(W_1,R_{1,\implyint},P)$ and 
	$\M_2^{\implyint}=(W_2,R_{2,\implyint},P)$ respectively such that for all $u_1,v_1,w_1\in W_1$ 
	and all $u_2,v_2,w_2\in W_2$,
	\begin{align}
	\label{exp1}			R_{1,\implyint}u_1 v_1 w_1\mbox{ iff } & R_1 u_1 w_1\mbox{ and } R_1 v_1 w_1\\
	\label{exp2}			R_{2,\implyint} u_2 v_2 w_2\mbox{ iff } & R_2 u_2 w_2 \mbox{ and } R_2 v_2 w_2
	\end{align}
	One can show \cite{Auc14c} that for all $\phi\in\langProtoConex$ and all $w_1\in W_1$, 
	$M_1,w_1\models\phi$ iff $M_1^{\implyint},w_1\entailR\phi$ (and likewise for $M_2$ and 
	$M_2^{\implyint}$). Now, a binary relation $\simproto$ between $M_1^{\implyint}$ and 
	$M_2^{\implyint}$ is a $\Conex$--bisimulation between $M_1^{\implyint}$ and 
	$M_2^{\implyint}$ iff for all $M,M'\in\{M_1^{\implyint},M_2^{\implyint}\}$,  all 
	$\x,\x',\y',\z'\in\Assign\cup\Assignb$ and all $p\in\Prop$, 
	\begin{itemize}
		\item  if $\x\simproto \x'$ and  $\x\in\exto{p}$ then $\x'\in\exto{p}$ (condition for $p$);
		\item  if $\y\simproto \y'$ and $R_{\implyint}' \y' \z' \x'$ then there are $\z,\x\in W$ such that $\z'\simproto \z$, $\x\simproto \x'$ and $R_\implyint \y \z \x$ $(*)$ (condition for $\implyint$);
		\item  conditions for $\top$ and $\bot$ trivially hold because of their semantics.
	\end{itemize}
	Using Expressions \eqref{exp1} and \eqref{exp2}, one can easily show that condition $(*)$ is equivalent to the following condition:
	\begin{itemize}
		\item if $\y\simproto \y'$ and $R' \y' \x'$ and $R' \z' \x'$ then there are $\z,\x\in W$ such that $\z'\simproto \z$, $\x\simproto \x'$ and $R \y \x$ and $R \z \x$ $(**)$.
	\end{itemize}
	We will show in Section \ref{sec:8} that  condition $(**)$  is equivalent on $\omega$--saturated models to Olkhovikov's condition ``step'' of \cite[Definition 1]{Olk17} of his ``basic asimulation''. 
\end{example}


\subsection{Molecular logics}

\renewcommand{\x}{w}
\renewcommand{\y}{v}
\renewcommand{\z}{u}

\begin{definition}[$\Conex$--bisimulation for molecular connectives]\label{def:invb}
	Let $\Conex$ be a set of molecular connectives and let $M_1=(W_1,\Rset_1)$ and $M_2=(W_2,\Rset_2)$ be two $\Conex$--models. 
	For all $\comc_0\in\Conex$, let  $V_{\comc_0}$ be the  vertices of the decomposition tree  $T_{\comc_0}$. We associate to each vertex $\comc\in V_{\comc_0}$  a   binary relation  $\simproto_\comc\subseteq\underset{\mult\in\mathbb{N}^*}{\bigcup}(W_1^\mult\times W_2^{\mult})\cup(W_2^{\mult}\times W_1^\mult)$. The set of such binary relations is  
	denoted  $\left\{\simproto\right\}\cup\underset{\comc_0\in\Conex}{\bigcup}\left\{\simproto_{\comc}\bmid \comc\in V_{\comc_0} \right\}$    
	and is  such that if $\comc$ is $\id_{\mult}$ for some $\mult\in\mathbb{N}^*$   
	then $\simproto_{\comc}$ is $\simproto$ and we  have that  $\simproto\subseteq\bigcap\left\{\simproto_{\comc}\bmid \comc\in \Conex \right\}$. We say that  this set of binary relations  is a \emph{$\Conex$--bisimulation} between $\M_1$ and $\M_2$  
	when  for all $\comc_0\in\Conex$,  all  vertices $\comc\in V_{\comc_0}$,  	if $\{\M,\M'\}=\{\M_1,\M_2\}$ then 
	for all  $\vec{\x}_1,\ldots,\vec{\x}_n,\vec{\x'}_1,\ldots,\vec{\x'}_n,\vec{\x},\vec{\x'}\in \Assign\cup\Assignb$, 
	\begin{enumerate}
		\item if  $\comc$ is an propositional letter $p$ then,   $\vec{\x}\simproto_\comc \vec{\x'}$ and $\vec{\x}\in\exto{p}$  imply  $\vec{\x'}\in\exto{p}$;
		\item if  $\comc$ has skeleton $\conn(\comc_1,\ldots,\comc_n)$  with  $\conn=(\sigma,\pm,\exists,\vecmult,(\pm_1,\ldots,\pm_n))$ and 
		we have   
		%
		$\vec{\x}\simproto_{\comc}\vec{\x'}$ and\linebreak $R^{\bb\sigma}_{\conn}\vec{\x}_1\ldots\vec{\x}_n\vec{\x}$, then 
		
		$\exists\vec{\x'}_1\vec{\x'}_2\ldots\vec{\x'}_n\left(\vec{\x}_1\pitchforkb_{\comc_1}\vec{\x'}_1\land\vec{\x}_2\pitchforkb_{\comc_2}\vec{\x'}_2\land\ldots\land\vec{\x}_n\pitchforkb_{\comc_n}\vec{\x'}_n\land R^{'\bb\sigma}_{\conn}\vec{\x'}_1\ldots\vec{\x'}_n\vec{\x'}\right)$;
		
		\item if  $\comc$ has skeleton $\conn(\comc_1,\ldots,\comc_n)$ with   $\conn=(\sigma,\pm,\forall,\vecmult,(\pm_1,\ldots,\pm_n))$ and 
		we have  
		%
		$\vec{\x}\simproto_{\comc}\vec{\x'}$ 
		
		and $-R^{'\bb\sigma}_{\conn}\vec{\x'}_1\ldots\vec{\x'}_n\vec{\x'}$, then 
		
		$\exists\vec{\x}_1\vec{\x}_2\ldots\vec{\x}_n\left(\vec{\x}_1\pitchforkb_{\comc_1}\vec{\x'}_1\land\vec{\x}_2\pitchforkb_{\comc_2}\vec{\x'}_2\land\ldots\land\vec{\x}_n\pitchforkb_{\comc_n}\vec{\x'}_n\land -R^{\bb\sigma}_{\conn}\vec{\x}_1\ldots\vec{\x}_n\vec{\x}\right)$; 		
	\end{enumerate} 
	where  for all   $j\in\exto{1;n}$, we have 	
	$\vec{\x}_j\pitchforkb_{\comc_j}\vec{\x'}_j\eqdef \begin{cases}
	\vec{\x}_j\simproto_{\comc_j}\vec{\x'}_j & \mbox{if $\pm_j=+$}\\
	\vec{\x'}_j\simproto_{\comc_j} \vec{\x}_j & \mbox{if $\pm_j=-$} 
	\end{cases}$. 

	When such a set of binary relations exists and is such that $\vec{\x}\simproto\vec{\x'}$, we say that $(M,\vec{\x})$ and $(M',\vec{\x'})$ are \emph{$\Conex$--bisimilar} and we write it  $(M,\vec{\x})\simul_{\Conex}(M',\vec{\x'})$.  
\end{definition}

Note that  case 1.\  is a  particular instance  of cases 2.\ and 3.\  with $n=0$. 

\begin{definition}\label{def:20}
	Let $\Conex$ be a set of molecular connectives. For all $\comc_0\in\Conex$ and all  vertex $\comc$ of the decomposition tree $T_{\comc_0}$, we define the language $\langProto_{\comc\Conex}$ as follows:
	\begin{align*}
	\langProto_{\comc\Conex}\eqdef\begin{cases}
	\left\{\comc(\phi_1,\ldots,\phi_n)\bmid \phi_1,\ldots,\phi_n\in\langProto_\Conex \right\} & \mbox{if $\comc$ is of arity $n>0$}\\
	\{p\}	 & \mbox{if $\comc=p$ is a propositional letter}\\
	\langProtoConex
	& \mbox{if $\comc$ is  $\id_{\mult}$ for some $\mult\in\mathbb{N}^*$.}
	\end{cases}
	\end{align*}
	Let $(M,\vec{\x})$ and $(M',\vec{\x'})$ be two pointed $\Conex$--models. We write $(M,\vec{\x})\equivLang_{\comc\Conex}(M',\vec{\x'})$ when for all $\phi\in\langProto_{\comc\Conex}$, $(M,\vec{\x})\entailR\phi$ implies  $(M',\vec{\x'})\entailR\phi$. We also write $(M,\vec{\x})\equivLang_{\Conex}(M',\vec{\x'})$ when for all $\phi\in\langProtoConex$, $(M,\vec{\x})\entailR\phi$ implies  $(M',\vec{\x'})\entailR\phi$. 
\end{definition}

\begin{proposition}\label{prop:9}
	Let $\Conex$ be a set of molecular connectives and  let $M_1=(W_1,\Rset_1)$ and $M_2=(W_2,\Rset_2)$ be two $\Conex$--models. 
	Let $\Conexo\subseteq\Conex$ and for all $\comc\in\Conexo$, let  $V_\comc$ be the  vertices of the decomposition tree  $T_\comc$. Let $\{\simproto\}\cup\underset{\comc_0\in\Conexo}{\bigcup}\left\{\simproto_\comc\bmid\comc\in V_{\comc_0}\right\}$ be a   $\Conexo$--bisimulation between $\M_1$ and $\M_2$.  If $\{M,M'\}=\{\M_1,M_2\}$  
	then for all $\comc_0\in\Conexo$ and all $\comc\in V_{\comc_0}$, for all $\vec{\x}\in\Assign$ and all $\vec{\x'}\in \Assignb$,
	if $\vec{\x}\simproto_\comc\vec{\x'}$ then  
	$(\M,\vec{\x})\equivLang_{\comc\Conexo}(\M',\vec{\x'})$. 
	In particular, 	if $\vec{\x}\simproto\vec{\x'}$ then  
	$(\M,\vec{\x})\equivLang_{\Conexo}(\M',\vec{\x'})$. 
\end{proposition}

\begin{definition}[Uniform connective]\label{def:21b}
	A \emph{uniform connective} is a molecular connective $\comc$ whose skeleton is of the form $\conn(\comc_1,\ldots,\comc_n)$ with $\conn=(\sigma,\pm,\fe,\vecmult,(\pm_1,\ldots,\pm_n))\in\Con$ such that 
	\begin{enumerate}
		\item  $n\geq 1$ and $\comc_1,\ldots,\comc_n$ are molecular skeletons  of arity 1; \item  for all $j\in\exto{1;n}$ such that $\comc_j\neq\id_{\mult}$ for all $\mult\in\mathbb{N}^*$, $\fe(\comc_j)=\begin{cases}\forall & \mbox{if $\pm_j=+$}\\ \exists & \mbox{if $\pm_j=-$}\end{cases}$;
		\item if $\comc_0$ is a molecular skeleton  appearing in the decomposition tree of $\comc$ of the form $\comc_0=\conn_0(\comc_1',\ldots,\comc_m')$  such that the tonicity signature of $\conn_0$ is $(\pm_1,\ldots,\pm_m)$,   
		then for all $i\in\exto{1;m}$,  $\fe(\comc_i')=\pm_i\fe(\comc_0)$.	\qedhere	
	\end{enumerate}
\end{definition}

According to our definition, molecular connectives of the form $\conn(\comc(\comc'(\comc'_1,\comc'_2)))$ cannot be uniform connectives, unless $\comc'_1$ or $\comc'_2$ is a propositional letter. This is due to our first condition: in that case,  $\comc(\comc'(\comc'_1,\comc'_2))$ should be of arity 1, which is possible only if $\comc'_1$ or $\comc'_2$ is a propositional letter. Hence, uniform connectives can be reduced to the composition of compound subconnectives  $\comc_i^1,\ldots,\comc_i^{m_i}$, each  of arity 1, so that molecular connectives are essentially of the form $\conn(\comc_1^1(\ldots \comc_1^{m_1-1}(\comc_1^{m_1})),\ldots,\comc_n^1(\ldots \comc_n^{m_n-1}(\comc_n^{m_n})))$. 
Basically, uniform connectives are such that the quantification patterns of their successive internal connectives   are essentially of the form $\exists\ldots\exists\ldots$ or $\forall\ldots\forall\ldots$ 

\begin{example}[Modal intuitionistic logic]\label{examp:9} Let $\Conex=\{p,\top,\bot,\land,\vee,\implyint,\conn,-\conn'\}$
	where $\conn,\conn'\in\Con^*$ are the molecular connectives  $\comc,\comc'$ of  Example \ref{example:9} and where $\{p,\top,\bot,\land,\vee,\implyint\}$ are defined in Example \ref{example:7}. 	The connectives of $\Conex$ are all uniform connectives. Note that $\conn'$ is not a uniform connective and that is why we consider $-\conn'$ for the moment, which is a uniform connective. We are going to see that we can easily get the bisimilarity condition for $\conn'$ from the bisimilarity condition for $-\conn'$. 
	
	Let $M_1=(W_1,\{R_1,R_{1,\Diamond},P\})$ and $M_2=(W_2,\{R_2,R_{2,\Diamond},P\})$ be two modal intuitionistic models. The set of binary relations $\{\simproto,\simproto_{\conn_2}^\conn,\simproto_{\conn_3}^{-\conn'}\}$ is a   $\Conex$--bisimulation iff for all $M,M'\in\{M_1,M_2\}$ with\linebreak $M=(W,\{R,R_\Diamond,P\})$ and $M'=(W',\{R',R'_\Diamond,P\})$, all $\x,\y,\z,\x',\y',\z'\in\Assign\cup\Assignb$ and all $p\in\Prop$, 
	\begin{itemize}
		\item  if $\x\simproto \x'$ and  $\x\in\exto{p}$ then $\x'\in\exto{p}$ (condition for $p$, like in  Example \ref{example:7});
		\item  if $\y\simproto \y'$ and $R' \y' \x'$ and $R' \z' \x'$ then there are $\z,\x\in W$ such that $\z'\simproto \z$, $\x\simproto \x'$ and $R \y \x$ and $R \z \x$ (condition for $\conn$, like in Example \ref{example:7});
		\item  if $R' \x' \y'$ and $\x\simproto \x'$ then there is $\y\in W$ such that $\y\simproto_{\conn_2}^\conn \y'$ and $R \x \y$,
		
		if $\x\simproto_{\conn_2}^\conn \x'$ and  $R'_\Diamond \x' \y'$  then there is $\y\in W$ such that $\y\simproto \y'$ and $R \x \y$
		
		(condition for $\conn=\conn_1(\conn_2)$);
		\item   if $R \x \y$ and $\x\simproto \x' $ then there is $\y'\in W'$ such that $\y'\simproto_{\conn_3}^{-\conn'} \y$ and $R' \x' \y'$,
		
		if $\x\simproto_{\conn_3}^{\-\conn'} \x'$ and  $R_\Diamond \x \y $  then there is $\y'\in W'$ such that $\y\simproto \y'$ and $R \x' \y'$
		
		(condition for $-\conn'=-\conn_1(\conn_3)$). 	
	\end{itemize}
	
	To obtain the bisimilarity condition for $\conn'$, it suffices to observe that for all $\comc\in\Con^*$, it holds that $(M,\x)\equivLang_{\{-c\}}(M',\x')$ iff $(M',\x')\equivLang_{\{c\}}(M,\x)$. So, we just have to replace $\x\simproto \x'$ by $\x'\simproto \x$ in the condition above. We obtain:
	\begin{itemize}
		\item[$(*)$] if $\x'\simproto \x$ and $R \x \y$ then there is $\y'\in W'$ such that $\y'\simproto_{\conn_3} \y$ and $R' \x' \y'$,
		
		if $\x\simproto_{\conn_3} \x'$ and $R_\Diamond \x \y$ then there is $\y'\in W'$ such that $\y\simproto \y'$ and  $R \x' \y'$.
	\end{itemize}
	
	It turns out that the	Conditions of $(*)$ are the conditions (diam--2(1)) and (diam--2(2)) of Olkhovikov \cite[Definition 9]{Olk17}, as expected. 
\end{example}

\section{Ultrafilters and  ultraproducts} 

\label{sec:5}

In that section, we  recall  some key notions  of model theory \cite{ChaKei98}, ultrafilters and  ultraproducts. 

\begin{definition}[Filter and ultrafilter] Let $I$ be a non--empty set. A \emph{filter $F$ over $I$} is a set $F\subseteq \mathcal{P}(I)$ such that 
		$I\in F$;
		if $X,Y\in F$ then $X\cap Y\in F$;
		if $X\in F$ and $X\subseteq Z\subseteq I$ then $Z\in F$.
	A filter is called \emph{proper} if it is distinct from $\mathcal{P}(I)$. An \emph{ultrafilter over $I$} is a proper filter $U$ such that for all $X\in\mathcal{P}(I)$, $X\in U$ iff $I-X\notin U$. 
\end{definition}

In the rest of this section,  $I$ is a non-empty set  and $U$ is  an ultrafilter over $I$.

\begin{definition}[Ultraproduct of sets]
	For each $i\in I$, let $W_i$ be a non-empty set. 
	For all $(w_i)_{i\in I},(v_i)_{i\in I}\in \prod_{i\in I}W_i$, we say that $(w_i)_{i\in I}$ and $(v_i)_{i\in I}$ are \emph{$U$-equivalent}, written $(w_i)_{i\in I}\sim_U (v_i)_{i\in I}$, if $\{i\in I\bmid w_i=v_i\}\in U$.  
	Note that  $\sim_U$ is an equivalence relation on $\prod_{i\in I}W_i$.  The equivalence class of $(w_i)_{i\in I}$ under $\sim_U$ is denoted  $\prod_U w_i\eqdef\left\{(v_i)_{i\in I}\in \prod_{i\in I}W_i\bmid (v_i)_{i\in I}\sim_U (w_i)_{i\in I}\right\}$. 
	The \emph{ultraproduct of $(W_i)_{i\in I}$ modulo $U$} is $\prod_U W_i\eqdef\left\{\prod_U w_i\bmid (w_i)_{i\in I}\in \prod_{i\in I} W_i\right\}$. 
\end{definition}


\begin{definition}[Ultraproduct] 	Let $(\M_i,\assignment_i)_{i\in I}$ be a family of  pointed structures. The \emph{ultraproduct}\linebreak  $\prod_U(\M_i,\assignment_i)$ is the pointed structure $\left(\prod_U\M_i,\prod_U\assignment_i\right)$ where $\prod_U\assignment_i:\var\rightarrow\prod_U W_i$ is the assignment such that for all $x\in\var$, $\left(\prod_U \assignment_i\right)(x)=\prod_U \assignment_i(x)$ and $\prod_U\M_i=(\W_U,\Rset_U)$ is defined as follows: 
	\begin{itemize}
		\item $W_U=\prod_U \W_i$;
		\item for all $n+1$--ary relations $R^i_\conn$ of $M_i$, 
		the $n+1$--ary relation  $\prod_U R_{\conn}\in\Rset_U$ is defined  for  all $\prod_U w^1_i,\ldots,$ $\prod_U w^{n+1}_i\in\W_U$ by $\prod_U R_{\conn} \prod_U w^1_i\ldots \prod_U w^{n+1}_i$ iff $\left\{i\in I\bmid R^i_\conn w^1_i\ldots w^{n+1}_i\right\}\in U$.  \qedhere
	\end{itemize}			
	
\end{definition}

\begin{definition}[Closure under  ultraproducts]
	Let $K$   be a  class of pointed structures. 	
%
	We say that $K$ is \emph{closed under ultraproducts} when for all non-empty sets $I$, if for all $i\in I$  $(M_i,\assignment_i)\in K$   then $\prod_U (M_i,\assignment_i)\in K$  for all ultraproducts $U$ over $I$.
\end{definition}

\renewcommand{\x}{x}
\renewcommand{\y}{y}
\renewcommand{\z}{z}

\section{A generic van Benthem characterization theorem}

\label{sec:5}

In this section we  generalize the van Benthem characterization theorem for modal logic \cite[Theorem~2.68]{BlaRijVen01} to molecular  logics. We first show how atomic and molecular  logics can be naturally embedded into first--order logic.

\begin{definition}[Translation from atomic and molecular  logics to \logicFO] \label{def:21}
	Let $\Conex$ be a set of atomic connectives. 
		
		\noindent \emph{Syntax}. For all $\mult\in\mathbb{N}^*$ and all $\vec{x}\in\left(\var\cup\cons\right)^\mult$, we define the mappings $ST_{\vec{x}}:\langProtoConex^\mult\rightarrow\langFOfree$, where $\langProtoConex^\mult$ is the set of formulas of $\langProtoConex$ of type $\mult$,  
		as follows: 
		for all atoms  $\letterp\in\Conex$,  all  $\conn\in\Conex$ of skeleton $(\sigma,\pm,\fe,(\mult,\mult_1,\ldots,\mult_n),$ $(\pm_1,\ldots,\pm_n))$ and all $\phi_1,\ldots,\phi_n\in\langProtoConex$,
		\begin{center}
			\begin{tabular}{rcl}
				$ST_{\vec{\x}}(p)$&$\eqdef$&$\PP(\vec{\x})$\\
				$ST_{\vec{\x}}(\phi\land_\mult\psi)$ & $\eqdef$ & $ST_{\vec{\x}}(\phi)\land ST_{\vec{\x}}(\psi)$\\
				$ST_{\vec{\x}}(\phi\vee_\mult\psi)$ & $\eqdef$ & $ST_{\vec{\x}}(\phi)\vee ST_{\vec{\x}}(\psi)$\\
				$ST_{\vec{\x}}(\conn(\phi_1,\ldots,\phi_n))$&$\eqdef$&$\fe\vec{\x}_1\ldots\vec{\x}_n\left(*_1 ST_{\vec{\x}_1}(\phi_1)\times\ldots\times *_n ST_{\vec{\x}_n}(\phi_n)\times       
				\RP_\conn^{\pm\sigma}\vec{\x}_1\ldots\vec{\x}_n\vec{\x}\right)$
			\end{tabular}
		\end{center}
		where $\vec{x}_1,\ldots,\vec{x}_n$ are tuples of free variables of size $k_1,\ldots,k_n$,
		
		$\times=\begin{cases}\land & \mbox{ if $\fe=\exists$}\\ \vee & \mbox{ if $\fe=\forall$}\end{cases}$ and for all $j\in\exto{1;n}$, $*_j=\begin{cases}\neg & \mbox{ if $\pm_j=-$}\\ \mbox{empty} & \mbox{ if $\pm_j=+$}\end{cases}$.  
		
		
		%
	
	
	\noindent \emph{Semantics}. 	
	Let $(\M,\vec{w})$ be a pointed $\Conex$--model of type $\mult$ with 
	$\vec{w}=(w_1,\ldots,w_\mult)$. Let $\vec{x}=(x_1,\ldots,x_\mult)\in\left(\var\cup\cons\right)^\mult$. A \emph{pointed 
		structure associated to} $(\M,\vec{w})$ \emph{and $\vec{x}$}  is a pointed structure $(\M,\assignment_{\vec{x}}^{\vec{w}})$ 
	(the set of predicates $\pred$ considered are 
	a copy of the relations of $\M$) where  the assignment $\assignment_{\vec{x}}^{\vec{w}}$ is such that  
	$\assignment_{\vec{x}}^{\vec{w}}(x_1)=w_1$,\ldots, $\assignment_{\vec{x}}^{\vec{w}}(x_\mult)=w_\mult$. 
	
	The above  translations canonically extend to molecular logics. Indeed, if $\Conex$ is a set of molecular connectives, every molecular formula of $\langProtoConex$ can be viewed as a formula of $\mathcal{L}_{\Conexb}$, where $\Conexb$ is the set of atomic connectives associated to $\Conex$. Likewise, any pointed $\Conex$-model can also be viewed as a pointed $\Conexb$-model. Then, we apply the above translations to obtain the translation of molecular formulas or $\Conex$-models into \logicFO. 
\end{definition}

The following proposition follows straightforwardly from the truth conditions of  Definition \ref{def:10}. 

\begin{proposition}\label{prop:7bd}
Let $\Conex$ be a set of molecular  connectives, let $(\M,\vec{w})$ be a pointed $\Conex$--model, let $\phi\in\langProtoConex$ of type $\mult$  and let $\vec{x}\in \var^\mult$. Then, $(\M,\vec{w})\entailR\phi$ iff $\left(\M,\assignment_{\vec{x}}^{\vec{w}}\right)\entailR ST_{\vec{x}}(\phi)$. 
\end{proposition}

\renewcommand{\x}{w}
\renewcommand{\y}{v}
\renewcommand{\z}{u}

\begin{theorem}[Characterization theorem]\label{theo:4} 
	Let $\Conex$ be a set of uniform  connectives complete for conjunction and disjunction. 	Let $\phi(\vec{x})\in\langFOfree$ with $\mult$   free variables $\vec{x}=(x_1,\ldots,x_\mult)$ 
	and   
	let $(\langage_{\Conex},\classConexE,\models)$ be a molecular  logic such that all models of $\classConexE$  contain relations $\left\{R_\conn\bmid \conn\in\Conex \right\}$  interpreting all the predicates 
	occuring in $\phi(\vec{x})$. Let $ST_{\vec{x}}(\classConexE)\eqdef\left\{(M,\assignment_{\vec{x}}^{\vec{w}})\bmid (M,\vec{w})\in\classConexE\mbox{ of type }\mult\right\}$ and assume that $ST_{\vec{x}}(\classConexE)$ is closed under ultraproducts. 
	The two following statements are equivalent:
	\begin{enumerate}
		\item There exists a formula $\psi\in\langage_{\Conex}$ such that $\phi(\vec{x})\leftrightarrow ST_{\vec{x}}(\psi)$ is valid on $ST_{\vec{x}}(\classConexE)$;
		\item \emph{$\phi(\vec{x})$ is invariant for  $\Conex$--bisimulations on $\classConexE$}, that is,  for all pointed  $\Conex$--models  $(\M,\vec{w}),  (M',\vec{w'})$ of $\classConexE$ of type $\mult$  
		such that 
		$(\M,\vec{w})\simul_{\Conex}\left(M',\vec{w'}\right)$,  we have that $\left(\M,\assignment_{\vec{x}}^{\vec{w}}\right)\models \phi(\vec{x})$ implies $\left(M',\assignment_{\vec{x}}^{\vec{w'}}\right)\models \phi(\vec{x})$.  
	\end{enumerate}
\end{theorem}

\begin{remark}
	The assumption that $ST_{\vec{x}}(\classConexE)$ is closed under ultraproducts is not really demanding since any class of structures definable by a set of first--order sentences is closed under ultraproducts by Keisler theorem \cite[Corollary~6.1.16]{ChaKei98}. For example, the class of (modal)  intuitionistic models is closed under ultraproducts since it is definable by a set of first--order sentences (we only need to impose the reflexivity and transitivity on the binary relations by means of the validity of corresponding sentences). So, our generic theorem applies to these logics. It also applies to modal logic and to many others since the class of all Kripke models is definable by an empty set of sentences and therefore is closed under ultraproducts. 
\end{remark}

\section{Related work}

\label{sec:8}

\subsection{Comparison with Olkhovikov's work}

The closest work to ours is by Olkhovikov \cite{Olk14,Olk17,Olk17b} who investigates generalizations of the van Benthem characerization theorm. The publications \cite{Olk14} and \cite{Olk17} deal in particular with (modal) intuitionistic (predicate) logic. Following our methodology, we have rediscovered  Olkhovikov's definitions in Examples \ref{example:7},  \ref{example:9} and \ref{examp:9}. In particular, we have the following result. 

\begin{fact}\label{fact:1}
	Let $M=(W,\{R,P\})$ and $M'=(W',\{R',P'\})$ be two $\omega$--saturated  intuitionistic models and let $\simproto$ be the maximal  $\Conex$--bisimulation between $M$ and $M'$ for set inclusion ($\Conex$ is defined in Example \ref{example:7}). Then, the following two conditions are equivalent:
	\begin{enumerate}
		\item Condition $(**)$ of Example \ref{example:7}: for all $v\in W$ and all $w',v',u'\in W'$,  if $v\simproto v'$ and $R' v' w'$ and $R' u' w'$ then there are $u,w\in W$ such that $u'\simproto u$, $w\simproto w'$ and $R v w$ and $R u w$;
		\item Condition ``step'' of \cite[Definition~1]{Olk17}: for all $v\in W$ and all $w',v'\in W'$, if $v\simproto v'$ and $R' v' w'$ then there is $w\in W$ such that $w\simproto w'$ and $w'\simproto w$ and $R v w$.
	\end{enumerate} 
\end{fact}

In \cite{Olk17}, Olkhovikov  also provides a generic van Benthem style characterization theorem  for a number of logics defined by specific kinds of connectives. He introduces  a normal form for connectives in terms of formulas of \logicFO\ that he calls \emph{$k$--ary guarded $x$--connectives}. A \emph{$k$--ary guarded $x$--connective of degree 0} $\mu=\psi(\PP_1(x),\ldots,\PP_k(x))$   is a Boolean combination of the unary predicates $\PP_1(x),\ldots,\PP_k(x)$. A \emph{$k$--ary $\forall$--guarded $x_1$--connective of degree $n+1$}  is a formula of the form $\forall x_2\ldots x_{m+1}\left(\bigwedge_{i=1}^m S_i(x_i,x_{i+1})\rightarrow \mu^-\right)$ where $S_1,\ldots,S_m$ are binary predicates and $\mu^-$ is a $k$--ary guarded $x_{m+1}$--connective of degree $n$ (provided that formula is not equivalent to $k$--ary guarded $x_1$--connective of a smaller degree). \emph{$k$--ary $\exists$--guarded $x_1$--connectives} are defined similarly. If one sets $\RP x_1\ldots x_m x_{m+1}$ for $\bigwedge_{i=1}^m S_i(x_i,x_{i+1})$ then $\forall x_2\ldots x_{m+1}$ $\left(\bigwedge_{i=1}^m S_i(x_i,x_{i+1})\rightarrow \mu^-\right)$ can be viewed as the first--order formula with free variable $x_1$ defining a molecular  connective: $\forall x_2\ldots x_{m+1}\left(x_2\in\exto{\bot}\vee\ldots\vee x_m\in\exto{\bot}\vee x_{m+1}\in\exto{\mu^-}\vee -\RP x_1\ldots x_m x_{m+1} \right)$. Hence, guarded connectives of degree not exceeding 1 are captured by specific molecular  connectives. It is unclear whether Olkhovikov's regular connectives of degree 2 are also captured by uniform connectives. 

In any case, our results strictly extend those of Olkhovikov because we are able to provide a van Benthem characterization for connectives defined by formulas of \logicFO\ with \emph{multiple} free variables. It is this feature that plays a key role for first-order logic \cite{Auc22}. 
 It is also made more clear and explicit  than in Olkhovikov's publication  how the suitable notions of bisimulation are defined from  logics given by their set of connectives. Finally, we showed in Examples \ref{example:9} and \ref{examp:9} how his results about (modal) intuitionistic logic \cite{Olk14,Olk17} can be recovered in our setting as specific instances of our general results. 

\subsection{Other related work}

Van Benthem  theorems 
have been  proved for many non--classical logics, such as 
(modal) intuitionistic logic \cite{Olk17}, 
intuitionistic predicate logic \cite{Olk14}, 
temporal logic \cite{KurRij97}, 
sabotage modal logic  \cite{AucEtAl18}, graded modal logic \cite{Rij00}, 
fuzzy modal logic \cite{WilEtAl18},
coalgebraic modal logics \cite{SchEtAl17}, 
neighbourhoud semantics of modal logic \cite{HanEtAl09}, 
the modal mu--calculus \cite{JanWal96},
hybrid logic \cite{AreEtAl01}. We showed that our generic Theorem \ref{theo:4} subsumes some of them \cite{Olk17,Olk14,KurRij97}. However, some others are not in the scope of our theorem because the correspondence language to which they refer  extends first--order logic. 
For example, the van Benthem  theorem for coalgebric modal logic \cite{SchEtAl17} is w.r.t.\  coalgebric predicate logic,  fuzzy modal logic \cite{WilEtAl18}  w.r.t.\  first--order fuzzy predicate logic and  the modal mu--calculus \cite{JanWal96}  w.r.t.\  monadic second--order logic.\\ 

\noindent \textbf{Acknowledgments.} I thank two anonymous reviewers for helpful comments.

\bibliographystyle{eptcs}
\bibliography{biblio.bib}

\end{document}